\documentclass[%
reprint,
superscriptaddress,
 amsmath,amssymb,
 aps,
prb,
]{revtex4-1}

\usepackage{graphicx}
\usepackage{dcolumn}
\usepackage{bm}

\usepackage{subfigure}
\usepackage{graphics,gensymb}
\usepackage{multirow}
\usepackage{rotating}
\usepackage{float}
\usepackage{appendix}
\usepackage{mathtools}
\usepackage{comment}
\usepackage{braket}
\usepackage{tabularx}
\usepackage{longtable}
\usepackage{libertine}
\usepackage{libertinust1math}
\usepackage{color}

\usepackage[bookmarks=false,hypertexnames=false]{hyperref}

\UndeclareTextCommand{\l}{OT1}
\DeclareTextSymbolDefault{\l}{T1}

\makeatletter
\let\NAT@bare@aux\NAT@bare
\def\NAT@bare#1(#2){%
	\begingroup\edef\x{\endgroup
		\unexpanded{\NAT@bare@aux#1}(\@firstofone#2)}\x}

\DeclareRobustCommand{\AA}{%
  \leavevmode
  \vbox{\ialign{##\cr
    \hidewidth\char'27 \hidewidth\cr
    \noalign{\nointerlineskip\kern-1.4ex}
    A\cr
  }}%
}
\makeatother

\newcommand{\quotemarks}[1]{``#1''}

\begin{document}

\preprint{APS/123-QED}

\title{Correlated local dipoles in PbTe}

\author{Boris Sangiorgio}
\affiliation{Department of Materials, ETH Z\"urich, Vladimir-Prelog-Weg 5, 8093 Z\"urich, Switzerland}
\author{Emil S.~Bozin}
\affiliation{Condensed Matter Physics and Materials Science Department, Brookhaven National Laboratory, Upton, NY 11973, USA}
\author{Christos D. Malliakas}
\affiliation{Department of Chemistry, Northwestern University, Evanston, IL 60208, USA}
\author{Michael Fechner}
\affiliation{Department of Materials, ETH Z\"urich, Vladimir-Prelog-Weg 5, 8093 Z\"urich, Switzerland}
\author{Arkadiy Simonov}
\affiliation{Department of Materials, ETH Z\"urich, Vladimir-Prelog-Weg 5, 8093 Z\"urich, Switzerland}
\author{Mercouri G.~Kanatzidis}
\affiliation{Department of Chemistry, Northwestern University, Evanston, IL 60208, USA}
\author{Simon J.~L.~Billinge}
\affiliation{Department of Applied Physics and Applied Mathematics, Columbia University, New York, NY  10027, USA}
\affiliation{Condensed Matter Physics and Materials Science Department, Brookhaven National Laboratory, Upton, NY 11973, USA}
\author{Nicola A.~Spaldin}
\affiliation{Department of Materials, ETH Z\"urich, Vladimir-Prelog-Weg 5, 8093 Z\"urich, Switzerland}
\author{Thomas Weber}
\affiliation{Department of Materials, ETH Z\"urich, Vladimir-Prelog-Weg 5, 8093 Z\"urich, Switzerland}

\date{\today}

\begin{abstract}
We present a combined single-crystal x-ray diffuse scattering and ab-initio molecular dynamics study of lead telluride, PbTe. Well-known
for its thermoelectric  and narrow-gap semiconducting properties, PbTe recently achieved further notoriety following the report of an unusual off-centering
of the lead atoms, accompanied by a local symmetry breaking, on heating. This observation, which was named \textit{emphanisis},  ignited considerable controversy regarding the details of the
underlying local structure and the appropriate interpretation of the total scattering experiments. In this study, we demonstrate close agreement between our diffuse scattering measurements and our calculations, which allows us to analyze features such
as higher-order correlations that are accessible in the simulations but not experimentally. This allowed us to discover an unusual correlated local dipole formation extending over several unit cells
with an associated local reduction of the cubic symmetry in both our x-ray diffuse scattering measurements and our molecular dynamics simulations.
Importantly, when averaged spatially or temporally, the most probable positions for the ions are at the centers of their coordination polyhedra. Our results therefore clarify the nature of the local symmetry breaking, and reveal the source of the earlier controversy regarding the existence or absence of off-centering. Finally,
 we provide an interpretation of the behavior in terms
of coupled soft optical and acoustic modes, which is linked also to the high thermoelectric performance of PbTe. 

\end{abstract}

\maketitle

Lead telluride (PbTe) is a narrow-gap semiconductor widely used in electronic and thermoelectric devices. 
Although its transverse optical (TO) phonon is soft, indicative of proximity to a ferroelectric phase transition\cite{Bate1970}, measurements of its bulk structure show that it retains its high-symmetry paraelectric $Fm\bar3m$ rocksalt structure at all temperatures. In contrast to its apparently simple average structural behavior, a recent analysis of the {\it local structure}, based on pair distribution function (PDF) analysis of neutron powder diffraction data, suggested the emergence of considerable local non-Gaussian distortions: The PDF peaks of the local structure broadened strongly with increasing temperature, accompanied by an asymmetry and the development of non-Gaussian lineshapes with shoulders. The PDF could be explained with a model that included off-centering of lead atoms with respect to their high symmetry positions, in the manner of fluctuating local dipoles, caused by Pb-Te dimerization, on heating to temperatures higher than 100 K. The phenomenon was called emphanisis \cite{bozin2010,Jensen2012}. 

The striking observation of an apparent local symmetry lowering {\it on warming}, in contrast to a conventional global symmetry-lowering phase transition that occurs {\it on cooling}, led to intense interest and many subsequent theoretical and experimental studies seeking to explain the behavior. A detailed understanding of the effect is still lacking, however, and indeed the existence of off-centering has even been questioned in some works. Significant anharmonicity was found in inelastic neutron scattering measurements~\cite{Delaire2011,Jensen2012}, consistent with the non-Gaussian PDF peaks, and the appearance of an additional phonon branch above 100~K in Ref.~\onlinecite{Jensen2012} was interpreted in terms of a dynamic local symmetry breaking on warming. However, while all experiments indicate large amplitude dynamic and anharmonic excursions of the Pb ions away from their central positions, there remains controversy regarding whether the probability distribution of the Pb ions is peaked at the high symmetry positions or displaced away from it on average. An analysis of synchrotron powder x-ray diffraction data using the maximum entropy method\cite{Kastbjerg2013} was consistent with large Pb atomic probability density away from the average position in all the lead chalcogenides, with refined magnitudes of the Pb displacement in PbTe (0.3~\AA\ at $300$ K) comparable to but exceeding that found in the original report\cite{bozin2010}.  Subsequent extended x-ray absorption fine structure (EXAFS) measurements\cite{Keiber2013a} reported that the Pb atomic probability distribution was peaked on average at the high symmetry position, albeit with large amplitude atomic displacement parameters (ADPs), and stated explicitly that the large displacements seen in diffraction measurements are inconsistent with the EXAFS data. A high-resolution neutron powder diffraction study found large increases in Pb ADPs on warming but did not reproduce the anomalies in their temperature dependence, attributing the earlier reports to errors in temperature calibration\cite{Knight2014}. Finally, a recent powder x-ray diffraction study that included the effects of Pb vacancies and microstrain in the Rietveld and maximum entropy method modeling\cite{Christensen2016} was consistent with a local symmetry breaking from cubic static off-centering of $0.2$ \AA\ or less at $100$ K. In addition, high concentrations of lead vacancies were found, pointing to sample dependence as a possible source of the inconsistencies between different measurements. 

On the theory side, DFT calculations in the original report~\cite{bozin2010} (using the local density approximation (LDA) and the self consistent \textit{ab initio} lattice dynamical (SCAILD) method~\cite{Souvatzis2009}) indicated a softer potential for local Pb displacements with increasing temperatures, counter to the usual hardening with increasing temperature of the free energy for a \textit{long-range} ferroelectric transition. The first \textit{ab initio} molecular dynamics (MD) study\cite{Zhang2011} (using the generalized gradient PW91 functional in a $216$-atom supercell within the VASP code) confirmed the soft and strongly anharmonic TO phonon but did not identify local dipole formation and concluded that the experimental results of~\citet{bozin2010} could be attributed to abnormally large-amplitude thermal vibrations. The calculated PDFs missed key physics, however, since while they showed broad peaks and some asymmetric character in the nearest neighbor Pb-Te peak, they did not reproduce the highly-non-Gaussian lineshapes and anomalous shifts in peak centroid with temperature seen in the data~\cite{bozin2010}.  A subsequent {\it ab initio} MD study\cite{Kim2012}, (again using the VASP package but this time with $64$-atom supercells), claimed to successfully reproduce the measured lattice dynamics, peak broadening with rising temperature, and non-Gaussian asymmetry and reported a larger lead off-centering than in the original experimental study. However, a comparison with the experimental data was not shown and the choice of exchange-correlation functional was  not stated, making it difficult to compare with the study of Ref.~\onlinecite{Zhang2011}. More recently, a novel slave-mode expansion method was used to calculate the finite-temperature dynamics of an $8000$-atom supercell\cite{Chen2014}. This study reproduced the observed splitting of the phonon peak on warming~\cite{Jensen2012}, and used the language of competing third- and fourth-order anharmonicities -- which tend to result in off-centerings -- rather than a local symmetry lowering to interpret the result. Classical MD simulations of $512$-atom supercells based on {\it ab initio} inter-atomic force constants (IFCs)\cite{Shiga2014} also reproduced the phonon splitting and identified unusually large cubic IFCs along the $\langle 100 \rangle$ direction as the source. In contrast, later studies\cite{Li2014,Li2014b} combining {\it ab initio} MD simulations (using $512$ atom supercells and the PBE functional in the VASP code) with single-crystal and powder neutron diffraction and inelastic neutron scattering suggested that a sharp resonance in the phonon self energy caused by nesting of phonon dispersions could combine with the anharmonicity to produce the large phonon splitting. Once again, however, while the simulated nearest-neighbor Pb-Te PDF peak showed increasing asymmetric character with temperature, the highly-non-Gaussian lineshapes and anomalous shifts in peak centroid were not captured. 

Finally, we note that similar behaviors and the associated controversies have also been reported in other materials. In related group IV chalcogenides, Ref.~\onlinecite{bozin2010} (supporting online material) already reported emergent local dipoles on warming in PbS, with the formation of shoulders on both sides of the PDF nearest neighbor peak, an interpretation supported in Ref.~\onlinecite{Kastbjerg2013}. More intriguingly, similar behavior has been reported in SnTe above its ferroelectric phase transition at low temperature \cite{Knox2014}, although again the temperature dependence has been disputed \cite{Knight2014}. The emergence of local displacements on warming is not exclusive to the group IV chalcogenides. In KNi$_2$Si$_2$, the average crystallographic structure is the same at all temperatures, but analysis of the local structure has been interpreted as an emergence on warming of local Ni displacements accompanied by differences in the electron density at the Ni sites, suggesting a fluctuating charge density wave\cite{Neilson2013}. In CsSnBr$_3$ a dynamic off-centering of Sn$^{2+}$ on warming has been reported, while again the average perovskite structure is not affected\cite{Fabini2016}.  Interestingly, in the latter case the PDF peaks showed similar asymmetries as in the earlier PbTe studies, but no clear shoulders. Finally, in La$_{1-x}$Ca$_x$MnO$_3$ local Jahn-Teller distortions have been reported in the high-temperature insulating phase that are not present in the low-temperature metallic phase, even though the average crystal structure is the same \cite{Billinge1996,Bozin2007}. 

In summary, some aspects of the originally reported emphanitic behavior of PbTe\cite{bozin2010,Jensen2012} are reproduced by all studies, notably the asymmetry of the PDF peaks and the splitting of the TO phonon. 
Other features, particularly the shoulders in the PDF peaks, remain elusive in the theoretical studies, thwarting a consistent interpretation of the behavior. In particular, two seemingly contradictory interpretations need to be reconciled in order to fully understand the underlying physics: The picture of local off-centering, which is based largely on diffraction data, compared with the picture of strong anharmonicity, which is based primarily on measurements of the dynamics.  

Here we present the first single-crystal x-ray diffuse scattering study of the temperature dependence of the local structure of PbTe, which we interpret with the aid of new \textit{ab-initio} molecular dynamics simulations. 
We start by discussing the average structure and find, as expected, that the rocksalt structure is retained at all studied temperatures, with displacive disorder associated with positional fluctuations. Next, we study the local structure with an analysis of the diffuse scattering. We obtain a peculiar decay of the atomic pair correlations which we are able to explain with the aid of higher-order correlations extracted from our MD simulations: We identify spontaneous displacements of Pb ions relative to their Te neighbors, resulting in correlated local dipoles that propagate for several unit cells. 
This is consistent with the earlier description of emphanitic behavior. We find, however, that in spite of these correlated displacements, the most probable position for each individual Pb ion is on average at the center of its coordinating Te polyhedron. However, our new finding that the local dipoles are correlated between unit cells resolves the apparent controversy by providing a picture that is consistent with both previously conflicting pictures: If at one instant in time, a Pb ion is displaced from the center of its coordination octahedron along $[100]$ say, the Pb ion in the neighboring unit cell in that direction is also displaced along $[100]$, and so on, over a correlation length of a few unit cells. The emergence of such fluctuating but correlated local dipoles on warming may then be considered as the definition of emphanisis. 
The correlations between the dipoles fall off in distance and in time and are evident in diffraction experiments only in the diffuse scattering signal, and propagates to the PDF in a total scattering experiment, and the 3D-$\Delta PDF$ measurements described here.  From our calculations we identify a possible origin for the asymmetry of the PDF peaks as an alternation of short and long bonds, and establish a link between the correlated dipoles and the high thermoelectric performance of PbTe.

\section{Average structure and lattice dynamics}
\label{sec:avg_structure}

We begin with an experimental and theoretical determination of the average atomic displacements as a function of temperature, focusing in particular on whether the atomic displacements are best approximated as single minimum harmonic or anharmonic oscillators or if the atoms occupy multi-minimum split positions.  
The experimental and computational details can be found in the Appendix. 
The refinement of the average crystal structure was done with the program \textit{SHELXL}\cite{Sheldrick2007} based on single crystal Bragg scattering. Atomic displacements were modeled in two approaches. 
First,  Pb and Te were located at the highest symmetry Wyckoff positions 4a (0,0,0) and 4b (0.5, 0.5, 0.5), and any static or dynamic displacements away from the high symmetry positions had to be captured by the harmonic atomic displacement parameters (ADPs) $U_{iso}^{\text{Pb}}$ and $U_{iso}^{\text{Te}}$ 
Second,  the atomic displacements were described as a convolution of discrete split vectors, which shift the atoms away from the high-symmetry to lower symmetry Wyckoff positions, and harmonic displacement functions. 
We tested such a split-position model only for Pb because its average displacements are larger than those of Te and refined models with split vectors $\langle x00\rangle$, $\langle xx0\rangle$ and $\langle 
xxx\rangle$. To avoid numerical correlations $U_{iso}^{\text{Pb}}$ and $U_{iso}^{\text{Te}}$ were constrained to the same values in each of the split model refinements. Consequently, each of our displacement models comprised two free structural parameters: $U_{iso}^{\text{Pb}}$ and $U_{iso}^{\text{Te}}$ in the harmonic model and a common $U_{iso}^{\text{Pb/Te}}$ plus a Pb split vector variable $x$ in the split models. The results of the refinements are summarized in Table \ref{table:compareModels}. 

\begingroup
\squeezetable
\begin{sidewaystable}
\caption{Overview of the results of the average structure refinements and of the MD simulations. Displacement parameters $U$ are given in \AA$^2$ units and the split parameter x is given in fractional coordinates. Note that the same values of $U^\text{Pb/Te}$ for the split models $\langle xx0\rangle$ and $\langle xxx\rangle$ are not a misprint but truly found in the refinements. The lattice constants reported are from the in-house measurements. The lattice constants observed in the synchrotron measurements are comparable. The displacement parameters from the MD simulations are calculated as mean square displacements of the atoms from the high-symmetric Wyckoff positions.}
\label{table:compareModels}
\begin{ruledtabular}
\begin{tabular}{cc|cccc|cccc|cccc|cccc|ccc}
\multicolumn{2}{c|}{} & \multicolumn{4}{c|}{Harmonic displacements} & \multicolumn{4}{c|}{Pb $\langle x00\rangle$ split model } & \multicolumn{4}{c|}{Pb $\langle xx0\rangle$ split model} & \multicolumn{4}{c|}{Pb $\langle xxx\rangle$ split model} & \multicolumn{3}{c}{MD}\\
T [K] & a [\AA] & $U_{iso}^{\text{Pb}}$ & $U_{iso}^{\text{Te}}$ & R1 & wR2  & $U_{iso}^{\text{Pb/Te}}$ & x(Pb) & R1 & wR2 & $U_{iso}^{\text{Pb/Te}}$ & x(Pb) & R1 & wR2 & $U_{iso}^{\text{Pb/Te}}$ & x(Pb) & R1 & wR2 & a [\AA] & $U_{iso}^{\text{Pb}}$ & $U_{iso}^{\text{Te}}$\tabularnewline\hline
293 & 6.4626(1) & 0.0260(2) & 0.0157(1) & 1.56 & 3.50 & 0.0153(2) & 0.0257(3) & 1.99 & 4.62 & 0.0150(1) & 0.0185(1) & 1.59 & 3.31 & 0.0150(2) & 0.0151(1) & 1.63 & 3.25 & 6.538 & 0.0182 & 0.0144\tabularnewline 
250 & 6.4546(1) & 0.0220(2) & 0.0130(1) & 1.53 & 3.49 & 0.0127(2) & 0.0244(3) & 1.75 & 4.13 & 0.0124(1) & 0.0175(1) & 1.46 & 3.06 & 0.0124(1) & 0.0143(1) & 1.51 & 3.00 & 6.532 & 0.0152 & 0.0120\tabularnewline 
200 & 6.4494(1) & 0.0182(2) & 0.0106(1) & 1.62 & 3.91 & 0.0104(2) & 0.0226(2) & 1.73 & 4.18 & 0.0102(1) & 0.0162(2) & 1.49 & 3.46 & 0.0102(1) & 0.0132(1) & 1.48 & 3.54 & 6.527 & 0.0123 & 0.0097\tabularnewline 
150 & 6.4420(1) & 0.0135(2) & 0.0077(1) & 1.64 & 3.93 & 0.0076(2) & 0.0199(2) & 1.72 & 4.28 & 0.0075(1) & 0.0141(1) & 1.65 & 3.96 & 0.0075(1) & 0.0115(1) & 1.68 & 4.05 & 6.521 & 0.0089 & 0.0070\tabularnewline 
125 & 6.4397(1) & 0.0115(2) & 0.0065(1) & 1.62 & 3.75 & 0.0064(2) & 0.0185(2) & 1.75 & 4.19 & 0.0063(1) & 0.0131(1) & 1.63 & 3.80 & 0.0063(1) & 0.0107(1) & 1.61 & 3.79 & - & - & - \tabularnewline 
\end{tabular}
\end{ruledtabular}
\end{sidewaystable}
\endgroup

The refinements did not provide a unique answer for the best average displacement model, however surprisingly, the $\langle x00\rangle$ split model, as was proposed by~\citet{bozin2010} and~\citet{Kastbjerg2013},  gave the highest R-values in all cases. All other models have very similar reliability factors, but  at high temperatures the $\langle xx0\rangle$ and $\langle xxx\rangle$ split models seem to be slightly better than the harmonic model. 

\begin{figure*} [hptb]
\includegraphics[width=.9\textwidth]{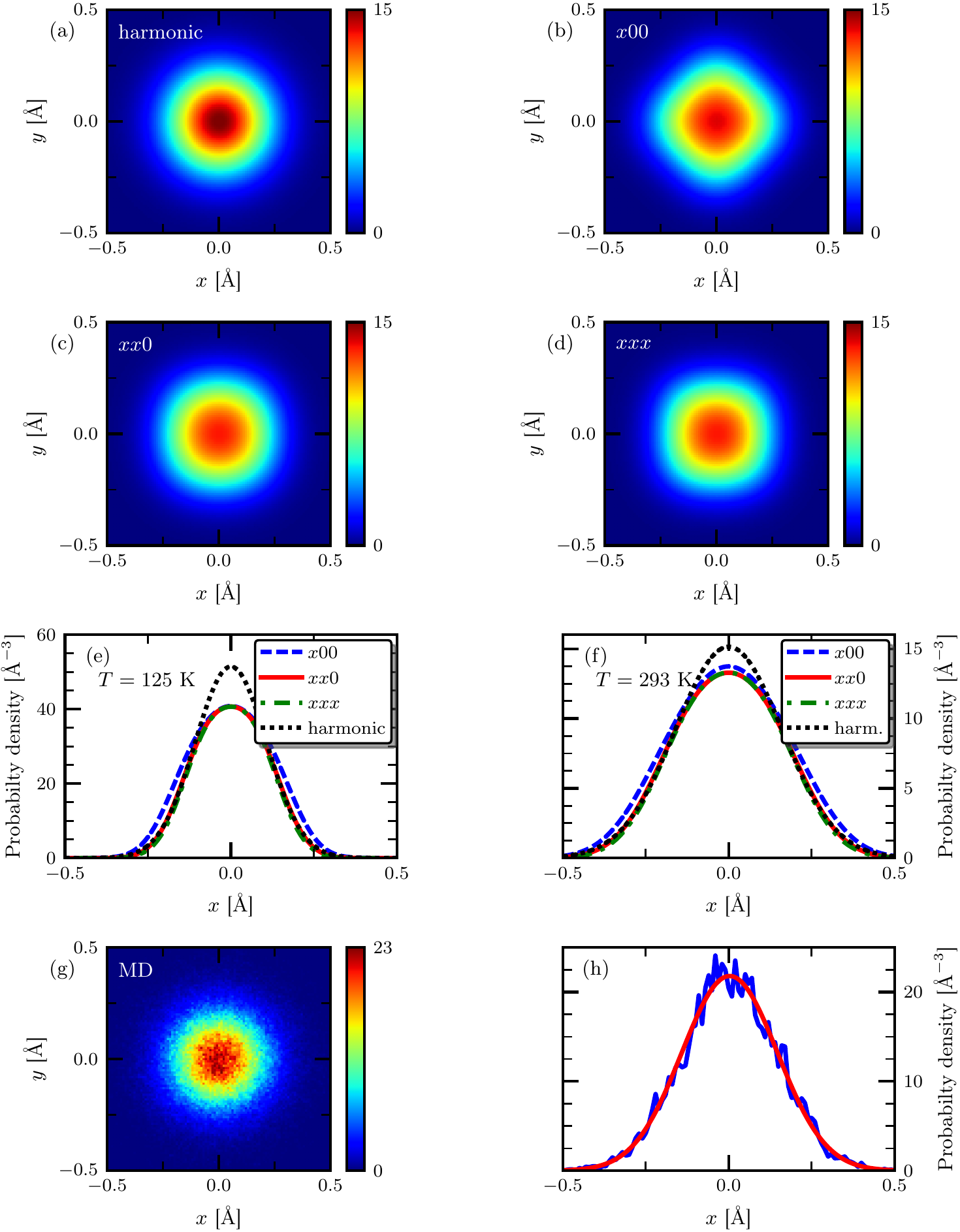}
\caption{Probability density functions of Pb. a) - d) show two-dimensional sections through the probability density functions at room temperature for the different split models indicated with the color scale representing the probability density in \AA$^{-3}$. e) and f) show scans through the center along the a-axis at $125$ K and at room temperature; (g) and (h) show the $2d$ section and scan through the center along the $a$ axis for our MD simulations both at $300$ K, where the red line in (h) represents a Gaussian fit.  }
\label{fig:probdens}
\end{figure*}

Figure~\ref{fig:probdens} shows the probability density functions obtained within the various displacement models. 
The $\langle x00\rangle$ split model represents the most anisotropic distribution function because it has the smallest number of split positions and needs to refine the largest Pb displacement per position to explain the non-thermal disorder, while the $\langle xx0\rangle$ and $\langle xxx\rangle$ split models appear more isotropic. 
It is important to note that even in the cases of the split models the convolution of the ADPs with the split vectors results in  a probability distribution function that is similar to a Gaussian. The major difference is that the tip of the distribution is slightly flattened and it is anisotropic, which suggests that Pb is located in a pseudo-harmonic, but slightly flattened potential. 
Our \textit{ab-initio} molecular dynamics simulations  (for details see Appendix) support this interpretation, though the resulting probability distribution (Fig.~\ref{fig:probdens}(g)) is more isotropic than that of any of the displaced models.  Our calculated average displacements (Figure \ref{fig:app-atom-displ} in the Appendix) are negligibly small for both Pb and Te atoms (consistent with the earlier molecular
dynamics simulations discussed above) and the probability density for the lead atoms (panel (g)) indicates that they are centered on the rocksalt high-symmetry positions. However, we do not have enough statistics to definitely assess whether the tip of the distribution is flattened (panel (h)).  

\begin{figure}[tb]
	\includegraphics[width=.95\columnwidth]{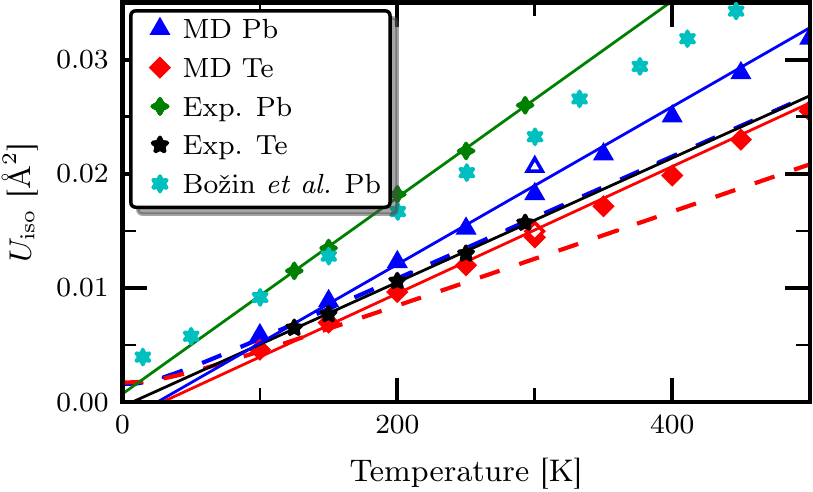}
	\caption{Atomic displacements parameters $U_{iso}$ obtained from our single crystal x-ray diffraction experiments (green for Pb and black for Te)  with the harmonic model and MD calculations (blue for Pb and red for Te). Light blue symbols are the temperature-corrected ADPs by \citet{Knight2014} of the original neutron diffraction based data by \citet{bozin2010}. Green and black (blue and red) continuous lines are a linear fit to the experimental (MD) data and show the  extrapolation to $0$ K.  
Blue and red dashed lines are ADPs computed in the harmonic approximation using the $T=0$ K phonons computed with DFT.
}
	\label{fig:adps}
\end{figure}

Figure \ref{fig:adps} shows the temperature evolution of the ADPs compared with the temperature-corrected data by \citet{Knight2014} of the original data of Ref.~\onlinecite{bozin2010}. 
Our refined ADPs show a linear dependence on temperature over the full examined temperature range, however with a larger slope than the  data of Ref.~\onlinecite{bozin2010}, which were collected with powder samples. 
Moreover, they nicely linearly extrapolate to close to zero at 0 K. 
We also extracted the ADPs from the MD simulations as the mean squared displacements from the average positions. They exhibit the same linear dependency and are consistently larger than those expected from a harmonic model using the $T=0$ K DFT phonons (computed using the Phonopy package\cite{phonopy}) pointing to anharmonic effects. While our calculated Te ADPs compare well with experiments, the Pb ADPs are severely underestimated in our MD simulations, due to their extreme sensitivity to the soft mode phonon frequencies and correspondingly to our supercell size (the open triangle at 300 K shows the result for a larger supercell) and neglect of spin-orbit coupling. 

In summary, we find good agreement in the temperature evolution of the average structure between our single-crystal Bragg measurements and \textit{ab-initio} MD simulations, as well as with the earlier literature reports; as expected the average structure is rocksalt. In particular, our calculated and measured ADPs are consistently larger than those calculated within a harmonic model. Our fits are unable to distinguish between split and single-minimum harmonic models, giving similar quality refinements in both cases. Therefore, while our average structure analysis certainly points to anharmonic behavior, it does not shed light on the question of the existence or nature of local off-centering. 

\section{Local structure}
\label{sec:local_structure}

We now analyze our diffuse scattering measurements and \textit{ab-initio} MD simulations to determine the local structure.

\subsection{Observations and qualitative interpretation of diffuse scattering and 3D-\texorpdfstring{$\Delta$}{Delta }PDF patterns}

\begin{figure*} [hptb]
    \includegraphics[width=.95\textwidth]{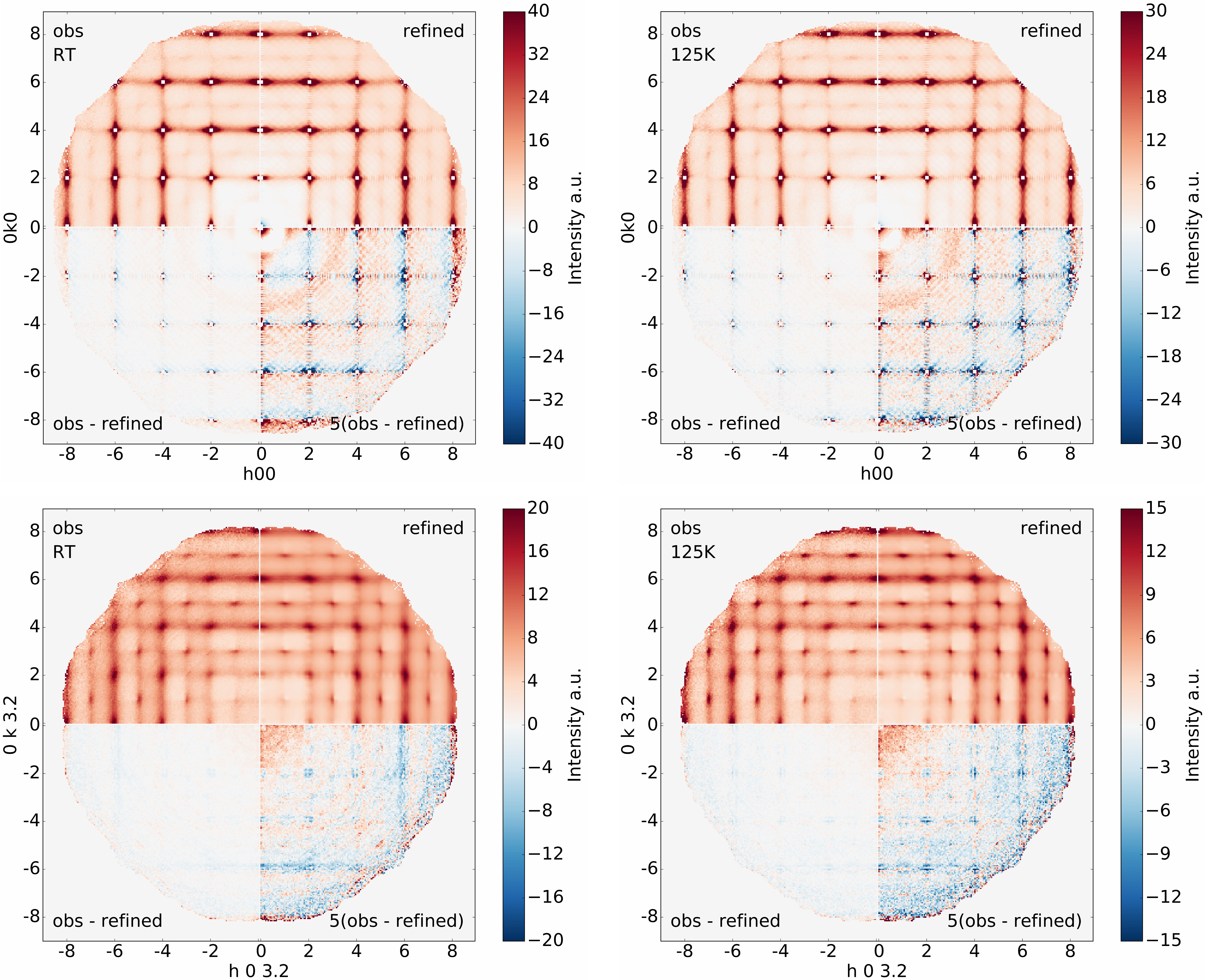}
\caption{Reciprocal space reconstructions of the diffuse scattering at room temperature (left panels) and 125 K (right panels). For an economic representation only the sections $hk$0 (upper panels) and $hk$3.2 (lower panels) are selected from the full data set having $360\times360\times360$ voxels. Observed intensities (\textit{obs}, upper left of each panel), results from the YELL refinement (\textit{refined}, upper right) and difference (lower) intensities of the diffuse scattering are compared. The patterns at $150$~K, $200$~K and $250$~K, which are not shown here, are comparable.  
The truncation ripples in the refined $hk0$ sections are due to incomplete coverage of the 3D-$\Delta$PDF maxima. 
The color wedges were linearly scaled by trial and error to allow a better qualitative comparison of the diffuse scattering patterns at different temperatures; the absolute diffuse intensities decrease with decreasing temperature as expected from the evolution of the ADPs. 
The white squares in the $hk0$ layers at $h,k = even$ are the cut-out Bragg reflection regions. Note that all quadrants include $h00$ and $0k0$ pixels or $h03.2$ and $0k3.2$ pixels, respectively, such that the pixels oriented up/down or left/right to the white lines separating the four sections in each panel have the same coordinates.}
\label{int_obs_calc}
\end{figure*}

The diffuse scattering pattern is dominated by a system of alternating weak and strong diffuse planes perpendicular to the cubic main axes. The planes are narrow but not sharp (Fig.~\ref{int_obs_calc}). 
The diffuse intensities are strongest beneath the Bragg reflections that are not extinct by the Bravais lattice. The overall intensities of the planes tend to increase with increasing diffraction vector, which is a characteristic pattern of displacive disorder with sub-\AA ngstrom amplitude. 
The weak diffuse intensities seen in the upper row in Fig.~\ref{int_obs_calc} can be understood as diffuse scattering from optical phonons, which scatter close to the Brillouin zone boundary. This is an interesting observation, because the typically small optical phonon amplitudes rarely allow measurements of corresponding diffuse scattering.    In addition, some structured band and square-like diffuse scattering is visible in higher layers (for an example see the lower row in Fig.~\ref{int_obs_calc}). 
It is interesting that the diffuse diffraction patterns show almost no variation within the investigated temperature range, apart from a decrease in the total diffuse intensity with decreasing temperatures. 
Therefore, it can be assumed that the qualitative local-order model is valid over the complete examined temperature range, $125 \le T \le 298$~K, with the amplitude of the displacements decreasing with decreasing temperature in this temperature range.

\begin{figure*} 
\centering
\subfigure{\includegraphics[width=0.7\textwidth]{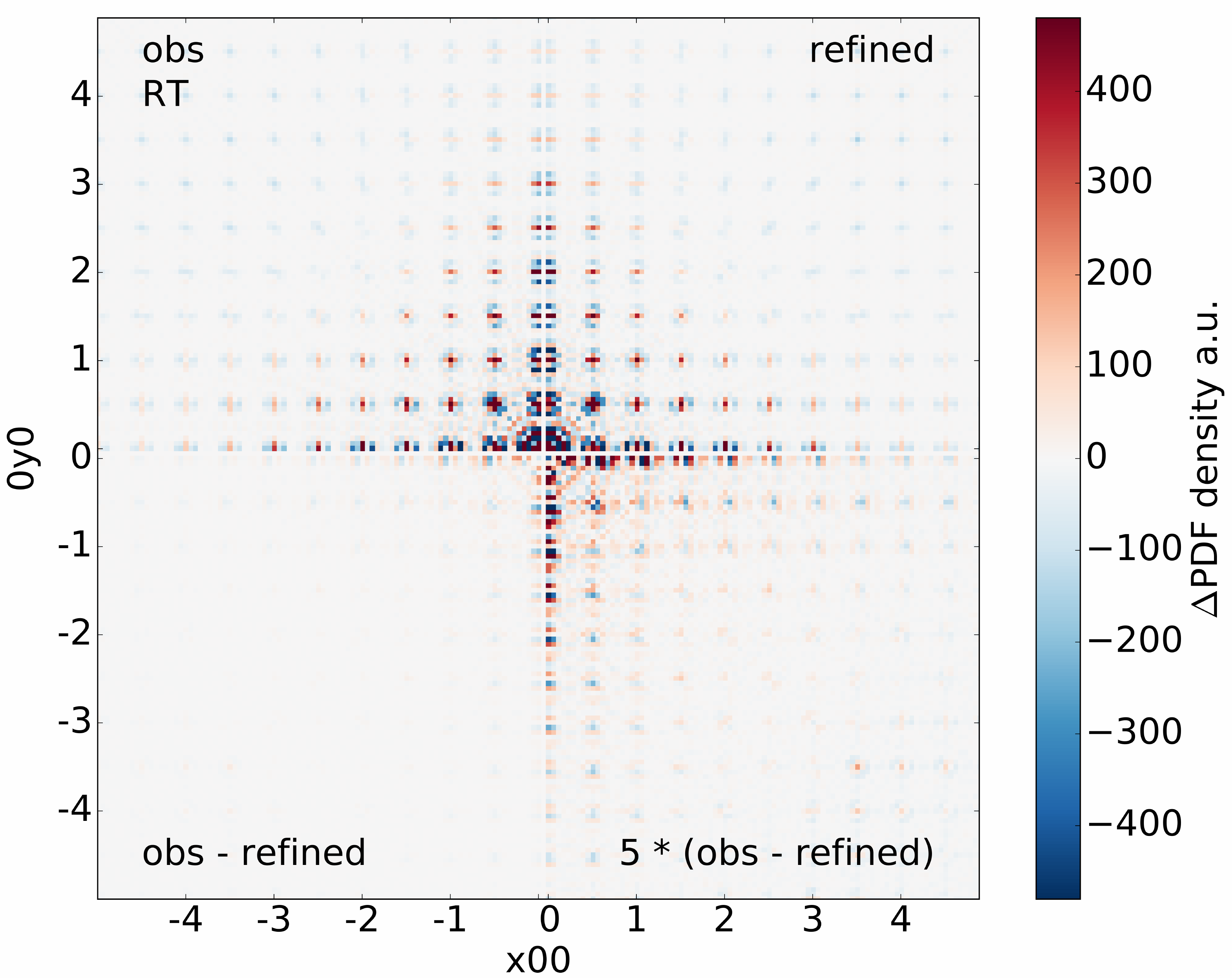}}\vfill
\subfigure{\includegraphics[width=0.7\textwidth]{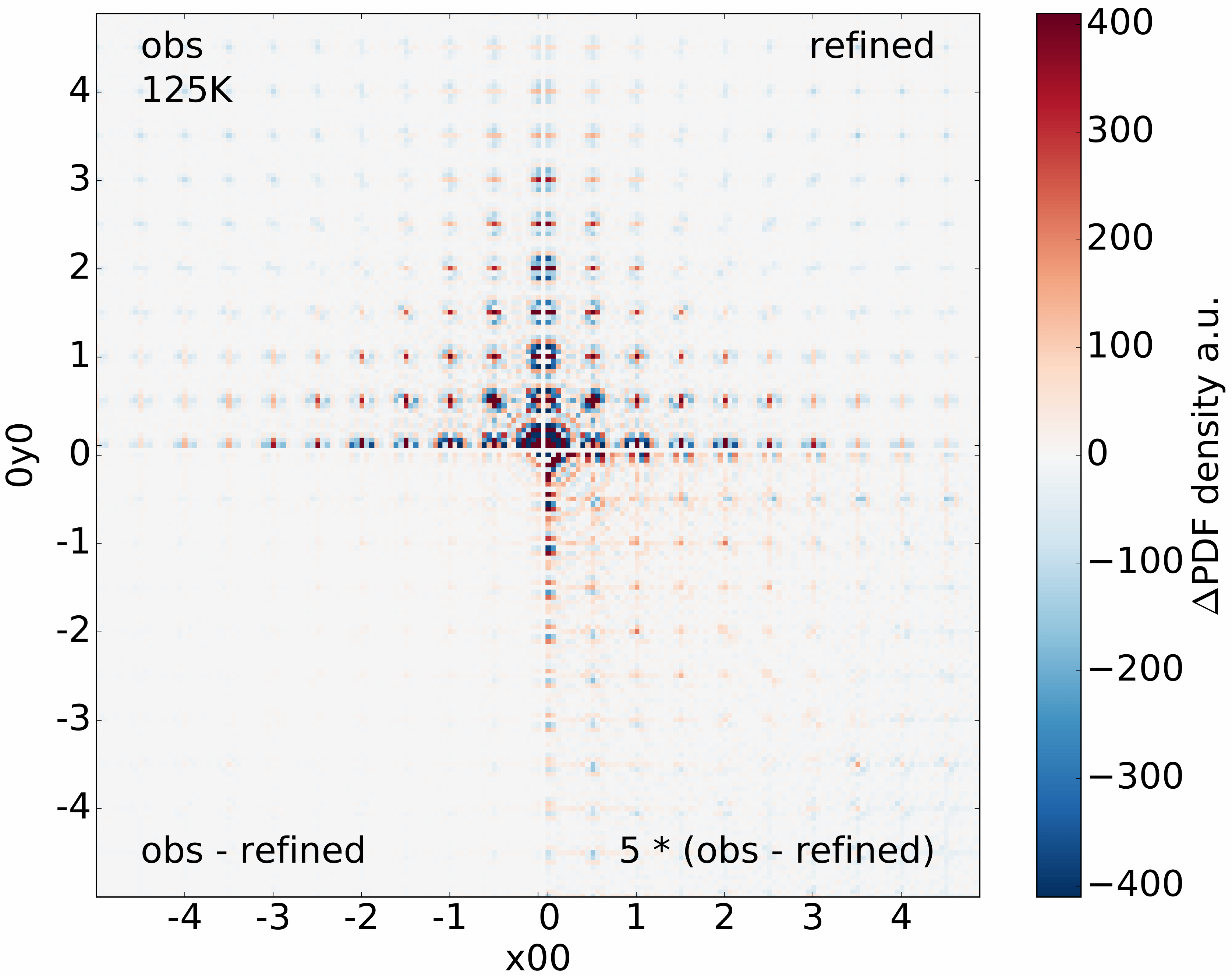}} 
\caption{3D-$\Delta$PDF $xy$0 layer at room temperature (top panel) and $125$~K (bottom panel). Observed densities (obs), results from the YELL refinement (refined) and difference 3D-$\Delta$PDF densities are compared. The patterns at $150$~K, $200$~K and $250$~K, which are not shown here, are comparable. The color wedges were linearly scaled by trial and error to allow a better qualitative comparison of the 3D-$\Delta$PDF densities at different temperatures. 
Note that absolute 3D-$\Delta$PDF densities decrease with decreasing temperature. Strongest disagreements between observed and refined densities are at $\langle x00\rangle$ coordinates. The origin may be anharmonic pair correlations that are not covered by our harmonic 3D-$\Delta$PDF model, but the resolution is not good enough to allow a clear conclusion.
The negative densities at large PDF distances come from artifacts due to masking diffuse scattering next to Bragg reflections. Note also that all quadrants include $x00$ and $0y0$ pixels, such that the pixels oriented up/down or left/right to the white lines separating the four sections in each panel have the same coordinates.}
\label{obs_calc_3DPDF}
\end{figure*}

Next, we focus on the interpretation of the 3D-$\Delta$PDF maps, obtained as the Fourier transform of the diffuse diffraction pattern, with the Bragg reflections cut out from the diffraction pattern (see Appendix for details). The 3D-$\Delta$PDF indicates where and how pair correlations of the real structure \textit{differ} from the average structure model as obtained from the Bragg reflections alone. Positive values mean that finding an atom at the end of the corresponding inter-atomic vector is more likely than in the space and time averaged structure, while the probability is lower if the 3D-$\Delta$PDF density is negative. For example, in the case that two atoms tend to move in-phase (\textit{positive correlation}) the corresponding 3D-PDF peak is narrower than in the Bragg scattering based 3D-PDF map. Thus the 3D-$\Delta$PDF peak shows positive values in the center and negative in the next neighborhood (looking like a Mexican hat pattern in the cross-section). If atoms move in anti-phase (\textit{negative correlation}) the behaviour is inverse (upside-down Mexican hat). At large distances 3D-$\Delta$PDF densities approach zero, because pair correlations of the real and the average structure become very similar due to the spatial loss of displacement correlations~\cite{Weber2012}. 
Figure \ref{obs_calc_3DPDF} shows the $xy0$ section of the 3D-$\Delta$PDF map at different temperatures. 
The 3D-$\Delta$PDF signals are found close to integer and half-integer lattice coordinates, corresponding to the average interatomic vectors of the rocksalt structure. 
Signals at $x + y + z = integer$ correspond to Pb/Pb and Te/Te interatomic vectors, which overlap perfectly, while those found at $x + y + z =$ \textit{half-integer} represent Pb/Te vectors. 
The PDF signals at overlapping homoatomic pairs are strongly dominated by local order properties of Pb/Pb pairs, because Pb has both a stronger scattering power and  larger ADPs, and therefore contributes more to the diffuse scattering and 3D-$\Delta$PDF maps. 
The homo- and hetero- interatomic vectors contribute similarly to the 3D-$\Delta$PDF, with the correlations strongest along $\langle 100\rangle$ directions and (see next section) weakest along $\langle 111\rangle$, although some minor differences can be observed after careful inspection. 
In general the 3D-$\Delta$PDF shows positive signals at average interatomic distances and negative signals along its radial neighborhood, which, as discussed above, is a typical signature for positive displacement correlations, i.e.\ it is consistent with a 3D-$\Delta$PDF pattern dominated by acoustic phonons.

\subsection{Pair correlations}
\label{sec:cor-coeffs-results}

\begin{figure}[hptb]
	\includegraphics[width=.85\columnwidth]{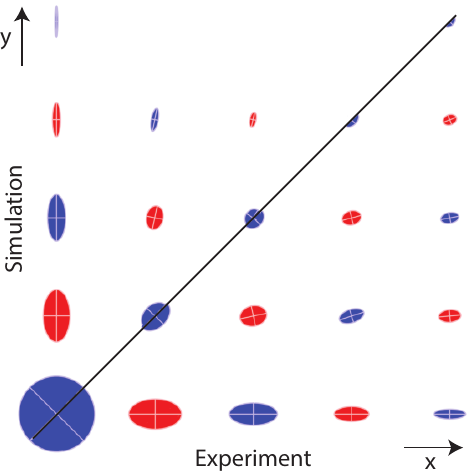}
	\caption{Graphical representation of the displacive pair correlations as obtained from the 3D-$\Delta$PDF refinement and from the MD simulations at RT and 300 K, respectively. The plot shows pair correlations in the $xy0$ layer. Blue ellipsoids represent correlations between homo-atomic pairs, while the red ellipsoids show correlations between Pb and Te. Sizes and orientations of the ellipsoids indicate the strength and direction of correlations, i.e.\ the strongest correlations found are along the main crystallographic axes.  The large blue circle in the bottom-left represents the correlation of an atom to itself, which by definition is unity. The diagonal line separates experimental and simulation results and coincides with a crystallographic mirror plane. All of the shown pairs are positively correlated along all directions. Missing ellipsoids in the simulated data have at least one, in all cases very weak, negative component indicating presence of some small negative correlations. Table \ref{table:pair-corrs-cf} in the Appendix presents an extended overview of the numerical values of the pair correlations.}
	\label{fig:corr-dft-2d}
\end{figure}

In this section we compare and analyze the pair correlations between atomic displacements as obtained from the 3D-$\Delta$PDF refinements and from the \textit{ab-initio} MD simulations. The displacive pair correlations were described in the harmonic approximation and refined with the program YELL\cite{Simonov2014}. For each symmetry independent average interatomic vector up to about 60 \AA\ we have refined the corresponding 3D covariance coefficients to mimimize the difference between the experimental diffuse scattering and the one calculated from the covariances. A detailed description of the 3D-$\Delta$PDF model and the refinement strategy is found in the Appendix. The PDF and reciprocal space results of the YELL refinements are shown in Figs.~\ref{int_obs_calc} and \ref{obs_calc_3DPDF}. The most significant pair correlations resulting from the YELL refinement are depicted in Fig.~\ref{fig:corr-dft-2d}, and compared to the results extracted from our MD simulations.  

\begin{figure*}[hptb]
	\includegraphics[width=.95\textwidth]{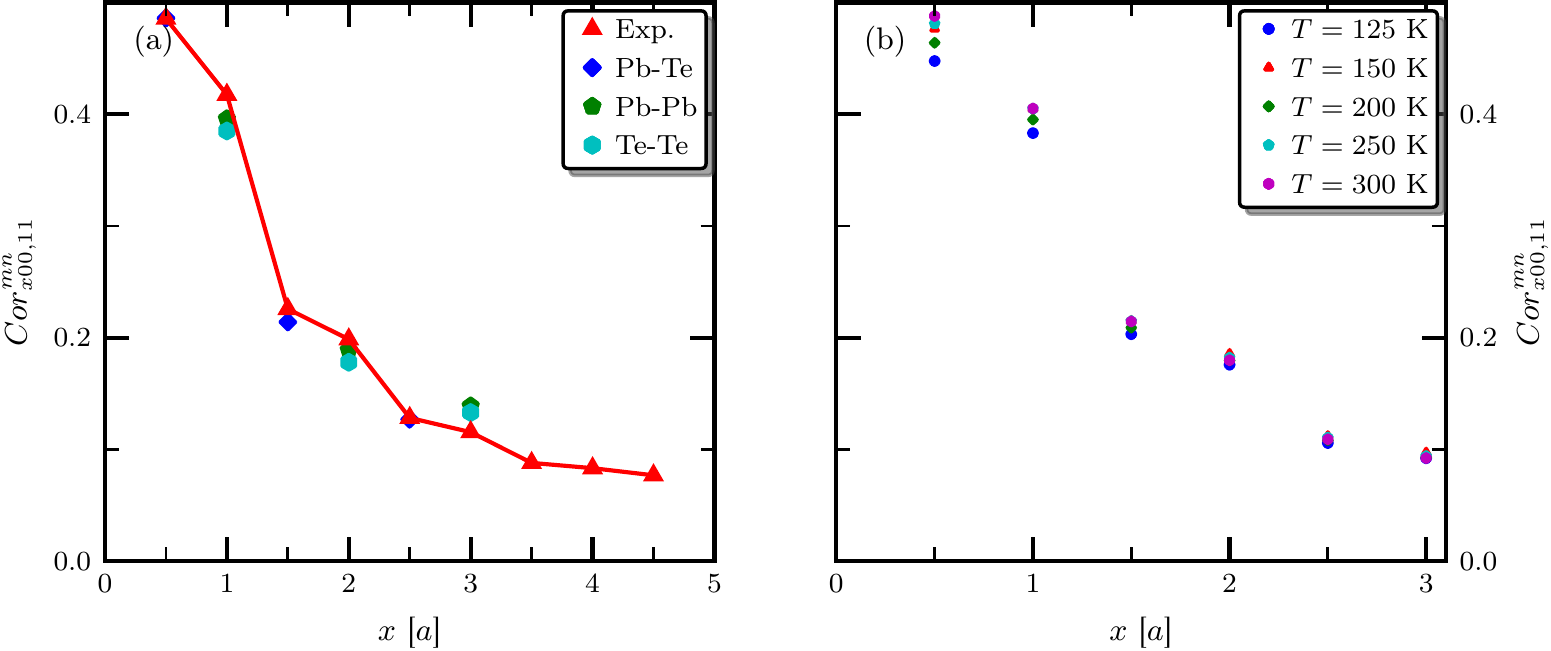}
	\caption{Longitudinal correlation coefficients along the cubic axis: (a) comparison at 300 K between diffuse-scattering extracted quantities (red triangles), and MD values [with the $6\times6\times6$ supercell]. The plateau and even small increase in the MD correlations for distances $2.5a$ and $3a$ is a consequence of periodic boundary conditions. (b) Temperature evolution of the experimental correlation coefficients. Whether the small temperature dependent spreads at a given distance are significant or if they are artefacts from incomplete models is not fully understood.} 
	\label{fig:corr-dft-1}
\end{figure*}

In agreement with the qualitative interpretation it is seen that the correlations are very strong for pairs separated by $\langle x00\rangle$ vectors with strong longitudinal correlations. Most importantly, pair correlations extracted from our MD match almost perfectly (see Fig.~\ref{fig:corr-dft-1}). 
In general the correlations decay quickly (Fig.~\ref{fig:corr-dft-1}), but with the formation of \quotemarks{steps} -- indicating that pairs of neighboring atoms have strongly correlated motions. Furthermore, correlations of the homo-atomic pairs are generally stronger than those of the hetero-atomic pairs. 
In agreement with our qualitative interpretation it is clear that the correlations are mostly independent of temperature over the temperature range studied. 
The amplitudes of the displacements change, however, in accordance with the changes in the ADPs. 

\begin{figure}[hptb]%
	\includegraphics[width=.95\columnwidth]{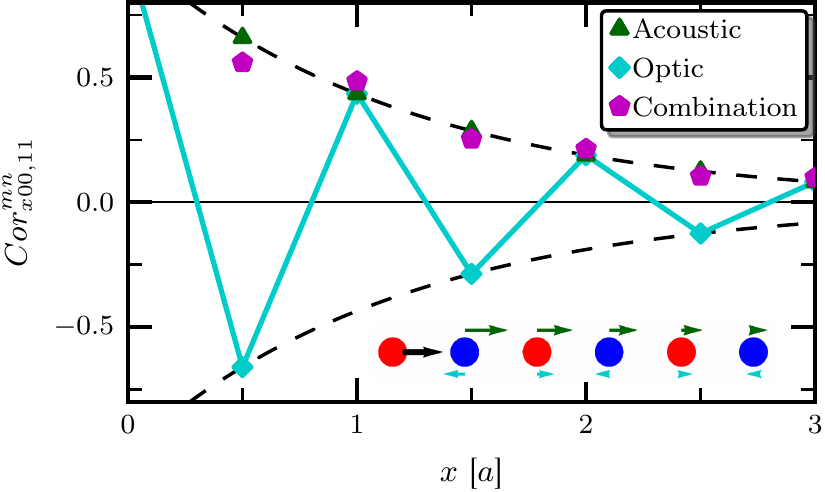}
	\caption{Cartoon of pair correlations expected for different \quotemarks{phonon modes} along the $\langle 100 \rangle$ direction; in green for acoustic-like displacements, in light-blue  for optic-like, and in purple for a superposition of the two. The effects are exaggerated to visualize the formation of steps. 
	}
	\label{fig:phonon-sketch}
\end{figure}

We now discuss a microscopic mechanism that is consistent with these pair correlations (see Fig.~\ref{fig:phonon-sketch}). 
As presented above, the 3D-$\Delta$PDF shows signatures of displacement correlations that are typical for acoustic phonons. 
These would generate positive pair correlations decaying with distance because of the short-range nature (green triangles in the figure). 
Optical phonon-like displacements, on the other hand, have pair correlations that alternate in sign (light blue diamonds). 
A superposition of both kinds of displacements, taking into account that in the real system the acoustic phonons dominate, gives rise to steps, with correlations of the homo-atomic pairs stronger than those of the hetero-atomic pairs. 
This displacement pattern suggests a Pb-Te dimerization along the $\langle 100 \rangle$ direction, that results in a local polarity. 

The presence of optical phonon-like displacements, which are usually not detected in diffuse scattering, implies not only that the optical phonons are active but also that their amplitude is big enough to be detected. We propose that this is possible in PbTe, because of the soft TO mode. 
To test this hypothesis we performed the same analysis using a literature Buckingham effective potential, with parameters fit to \textit{ab initio} calculations for PbTe \cite{Qiu2008,Qiu2012}. 
This potential was shown to reproduce reasonably the mechanical and phonon properties of bulk PbTe, except that the optical phonons calculated with the Buckingham potential are much harder than those in DFT. In particular, the TO mode at $\Gamma$ is in the order of $3$ THz instead of about $1$ THz in DFT. 
\begin{figure}[hptb]
	\includegraphics[width=.95\columnwidth]{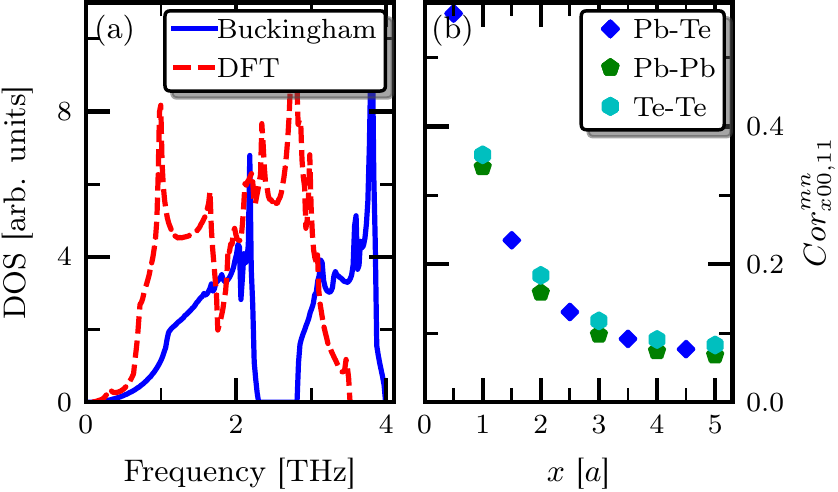}
	\caption{Properties derived from the Buckingham potential of Refs.~\onlinecite{Qiu2008,Qiu2012}. (a) Phonon density of states (DOS) [computed using the Phonopy package \cite{phonopy}]; in blue with the Buckingham potential, in dashed red with CP2K. 
    (b) Longitudinal correlation coefficients along the cubic axis at 300 K [with the $10\times10\times10$ supercell] computed using the Buckingham potential. }
	\label{fig:buck_corrs}
\end{figure}
Moreover, the optical and acoustic phonons are clearly separated in energy (Fig.~\ref{fig:buck_corrs}(a)), so that a coupling between them is unlikely. 
We performed classical molecular dynamics with this potential as implemented in LAMMPS \cite{Lammps} using a $10\times10\times10$ supercell [we checked  the results also with the $6\times6\times6$ supercell to make sure that the results in this supercell are not affected by finite-size effects] and, when possible, the same settings as we used in our ab-initio MD calculations. 
As expected, the pair correlations are found to decay smoothly without the formation of any steps  (Fig.~\ref{fig:buck_corrs}(b)).  

\begin{figure}[hptb]
	\includegraphics[width=.95\columnwidth]{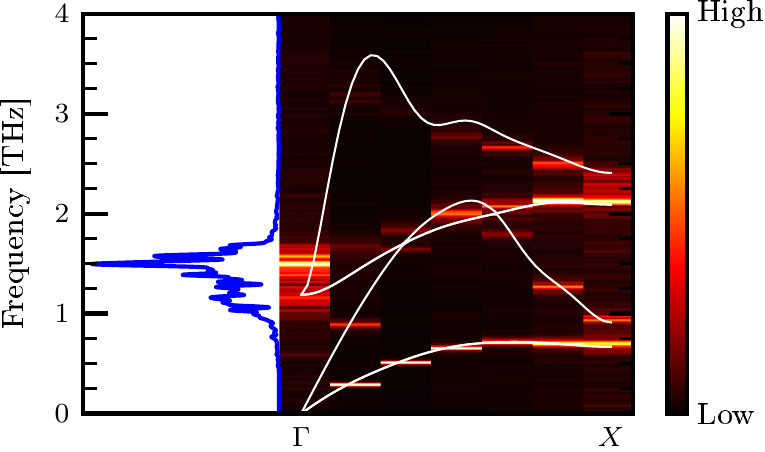}
	\caption{Power spectrum $Z(\mathbf{q},\nu)$ at $300$ K computed from the $6\times6\times6$ supercell using ab-initio MD. The left panel shows $Z(\mathbf{q},\nu)$ at the $\Gamma$ point (arbitrary units are used). The right panel shows the dispersion relation along the $\Delta$-line. White lines correspond to $T=0$ K DFT calculations.}
	\label{fig:phonons-MD}
\end{figure}

Interestingly, inelastic neutron scattering experiments by \citet{Delaire2011} reported a strong coupling between the ferroelectric TO mode and the longitudinal acoustic (LA) modes. 
One of its signatures -- the avoided crossing between TO and LA -- is also captured by our MD simulations. 
Figure \ref{fig:phonons-MD} presents the power spectrum $Z(\mathbf{q},\nu)$ at $300$ K computed as described in Ref.\ \onlinecite{Chen2014}, that can be directly compared to the inelastic neutron scattering intensities.  
The right panel shows the dispersion relation along the $\Delta$ direction; although the $T=0$ K DFT bands (white lines) have an LA-TO crossing, the MD results show a repulsion between LA and TO leading only to a contact between the two bands roughly midway along the $\Gamma$ to $X$ line. The left panel presents the power spectrum at the $\Gamma$ point. The peak is very broad and a splitting starts developing. Since these features are already highly enhanced with respect to a $3\times3\times3$ supercell, we expect that an even larger supercell would allow a good description of the emergence of the additional phonon mode, as in previous experiments and (classical) MD simulations~\cite{Jensen2012,Shiga2014,Chen2014,Li2014}. 

\begin{figure}[hptb]
	\includegraphics[width=.95\columnwidth]{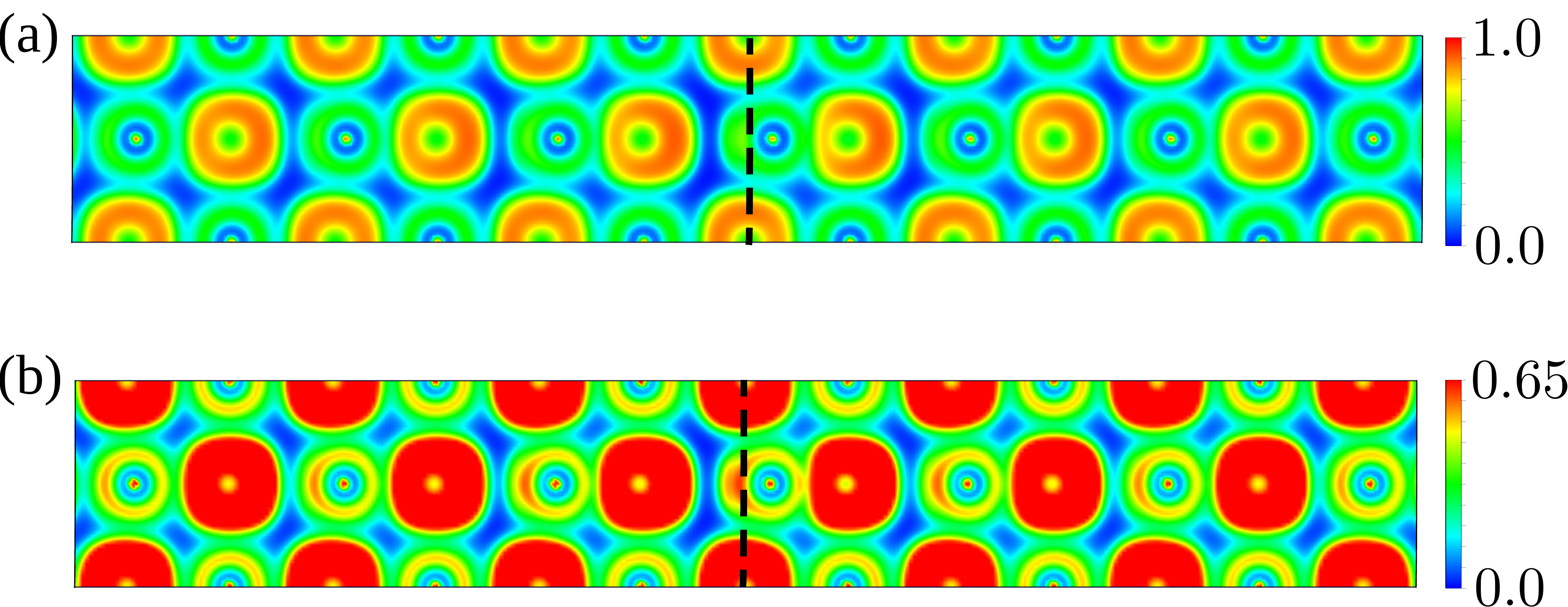}
	\caption{
Electron localization function (ELF) computed with the VASP~\cite{kresse1996} code for a structure distorted according to the pattern of pair correlations shown in Fig.~\ref{fig:corr-dft-1} with the central Pb ion in the $15\times1\times1$ supercell given the largest distortion from its high symmetry position (vertical dashed line). The displacements towards the right of the figure are exaggerated to amplify the effects. Different magnitude ranges are chosen for the upper and lower panels to illustrate the increased Pb-Te bonding as the Pb moves right towards its neighboring Te (a) and the formation of the lone pair (small red lobe) left of the Pb ions (b). 
    }
	\label{fig:elf}
\end{figure}

Finally for this section, we show the effect of the displacement pattern linked with the pair correlations on the electronic structure, by plotting in Fig.~\ref{fig:elf} the calculated electron localization function (ELF) for a static configuration consistent with the inset at the lower right of Fig.~\ref{fig:phonon-sketch}. The ELF provides a measure of the valence charge density, weighted to emphasize regions of higher electron localization, and has been used effectively to analyze the electronic structure in the related ferroelectric IV-VI chalogenides\cite{Waghmare2003}, where their stereochemical activity drives the dipole formation in the ferroelectric state. The central Pb atom is the most strongly displaced from its high-symmetry position (indicated by the black dashed line) with the distortion amplitudes gradually reducing with distance from the center.
In the top panel (a) the scale is chosen to illustrate the enhancement of the Pb-Te chemical bond as the Pb ions displace to the right towards their Te neighbors (at the centers of the orange regions). In the lower panel (b) the scale is chosen to emphasize the red ``banana-shaped'' localized electrons to the left of the displacing Pb ions, which form the sterochemically active lone pair. The stereochemical activity of this lone pair of electrons drives the displacement of the Pb ions and prevents the Te ions from acquiring a similar displacement leading to dipole formation.

Note that our calculations are consistent with the reported long-ranged interactions along $\langle 100 \rangle$ through resonant bonding \cite{Lee2014}  (or equivalently lone pairs) and the recent report of anisotropic microstrain along $\langle 100 \rangle$ in all lead chalcogenide systems by \citet{Christensen2016}. 
The latter paper argued that microstrain is a manifestation of the structural transition to an orthorhombic phase under pressure (in fact, the major atomic displacements in that phase transition are along $\langle 100 \rangle$\cite{Rousse2005}), based on what they call \quotemarks{anion-mediated Pb($6s$)-Pb($6p$) interaction}, another term for the (revised) lone pair\cite{walsh2011} or resonant bonding\cite{Lee2014} concept.

\subsection{Higher-order correlations}
\label{sec:many-body-corrs}

\begin{figure}[hptb]%
	\includegraphics[width=.8\columnwidth]{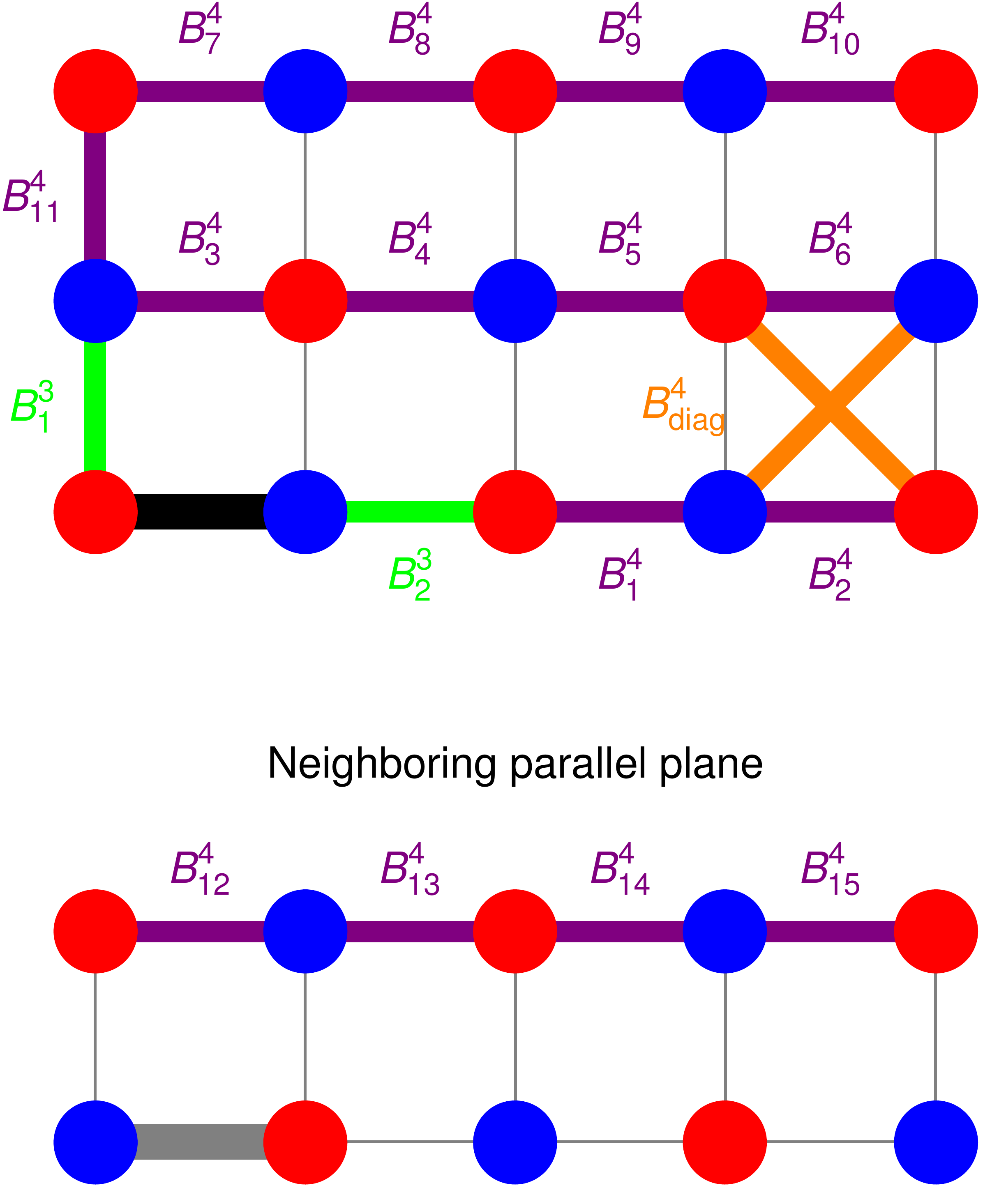}
	\caption{Notation of our computed higher-order correlations between pairs of nearby bonds. The reference bond (Bond 1 in the probability-density plots) is marked in black, with three-body correlations in green, four-body in purple, and correlations between diagonal bonds in orange. The lower panel shows a crystal plane lying above or below the upper panel (with the gray bond lying just on top or below the reference black bond). The upper index in our notation for higher-order correlations indicates the type of correlation (three- or four-body), and the lower enumerates them. 
	}
	\label{fig:corrs-sketch}
\end{figure}

The correlation of bond lengths provides valuable information about the local dipoles present in the structure. However, bond-length correlations represent a many-body problem, which cannot be easily extracted from Bragg and diffuse scattering alone \cite{Welberry1991}. To exceed this limitation we analyze the bond-lengths correlations from our MD simulations.  
Since we have shown that our MD reproduces the experimental pair correlations well, we are confident that our calculated higher-order correlations are physically relevant. 
Fig.~\ref{fig:corrs-sketch} shows all computed correlations. 
To quantitatively assess the kind of correlation present between each pair of bonds we compared to a reference state in which the bonds are uncorrelated (see Fig.~\ref{fig:app-corrs4-neut} in the Appendix); the latter was computed by considering all pairs of bonds at least $15$ \AA\ apart. 

\begin{figure*}[hptb]%
	\includegraphics[width=.95\textwidth]{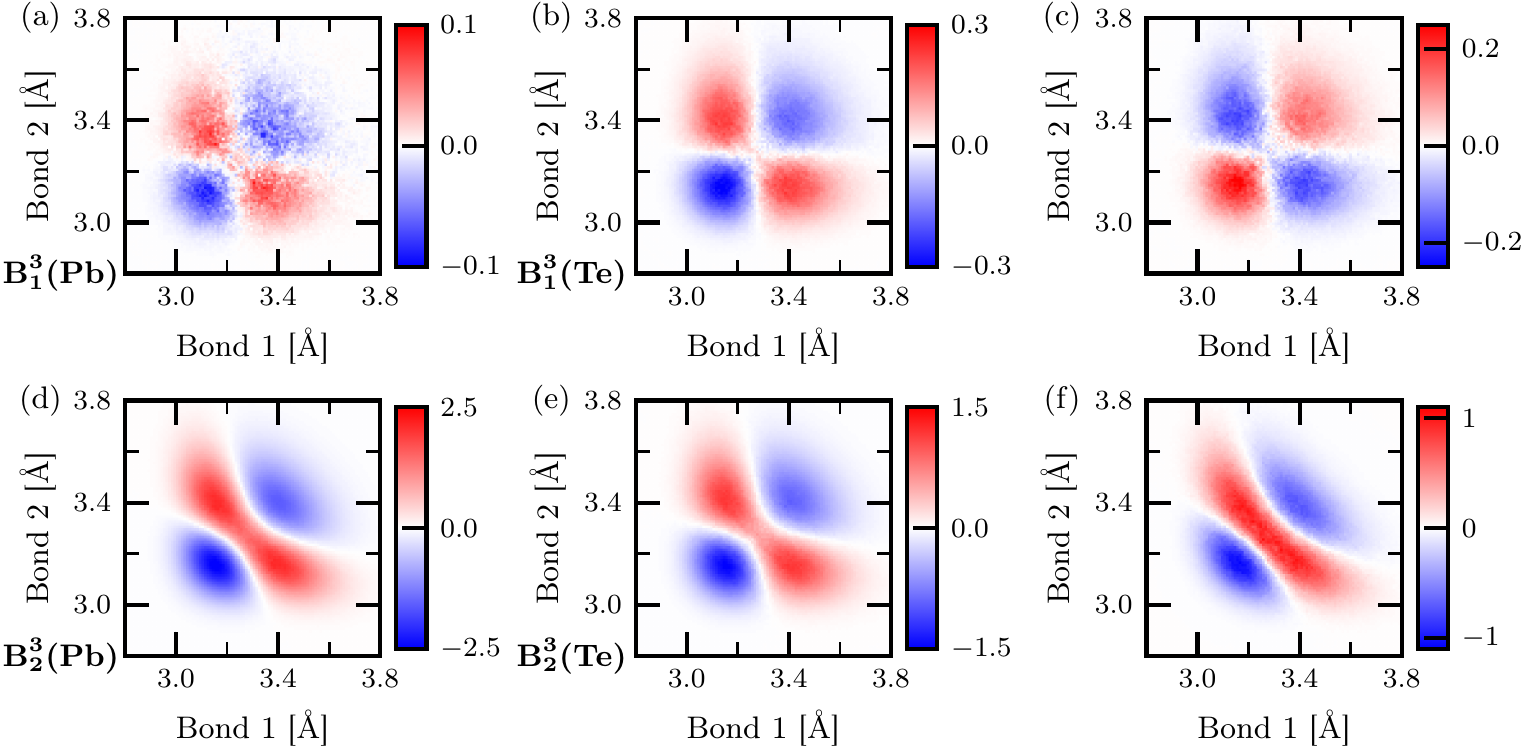}
	\caption{Difference probability densities of three-body correlations calculated from MD. The upper panel presents bonds $B^3_1$; the lower panel bonds $B^3_2$. 
    The first and second columns show difference densities with respect to the reference when the shared atom is Pb and Te, respectively, while the third column shows the difference between the first and second column. 
    The color scale represents the probability density in \AA$^{-2}$.
    }
	\label{fig:corrs3}
\end{figure*}

We start by showing the three-body correlations, i.e.\ the correlations between bonds that share an atom (Fig.~\ref{fig:corrs3}; for our notation see Fig.~\ref{fig:corrs-sketch}). In general, the correlations should depend on the type of shared atom, so we present our results for both Pb and Te shared. 
Fig.~\ref{fig:corrs3} shows probability densities for three-body correlations. Bonds $B^3_2$ (lower panel) show a clear negative correlation -- if one bond is longer than on average the other tends to be shorter. 
Note also that when the shared atom is Pb (d) the distribution is more elongated than for Te (e) -- as already discussed Pb is more polarizable, leading to a larger variability in the bond distribution. 
In the same way, $B^3_1$ bonds (upper panel) show a negative correlation, even though the deviations from the reference state are smaller. However, here larger deviations are observed for bonds whose shared atom is Te. 

We next analyze the further apart bonds shown in Fig.~\ref{fig:corrs4} in the Appendix. 
By close inspection of the four-body correlations' probability densities, the following local picture can be derived (Fig.~\ref{fig:dips-picture}): 
\begin{figure}[hptb]%
	\includegraphics[width=.95\columnwidth]{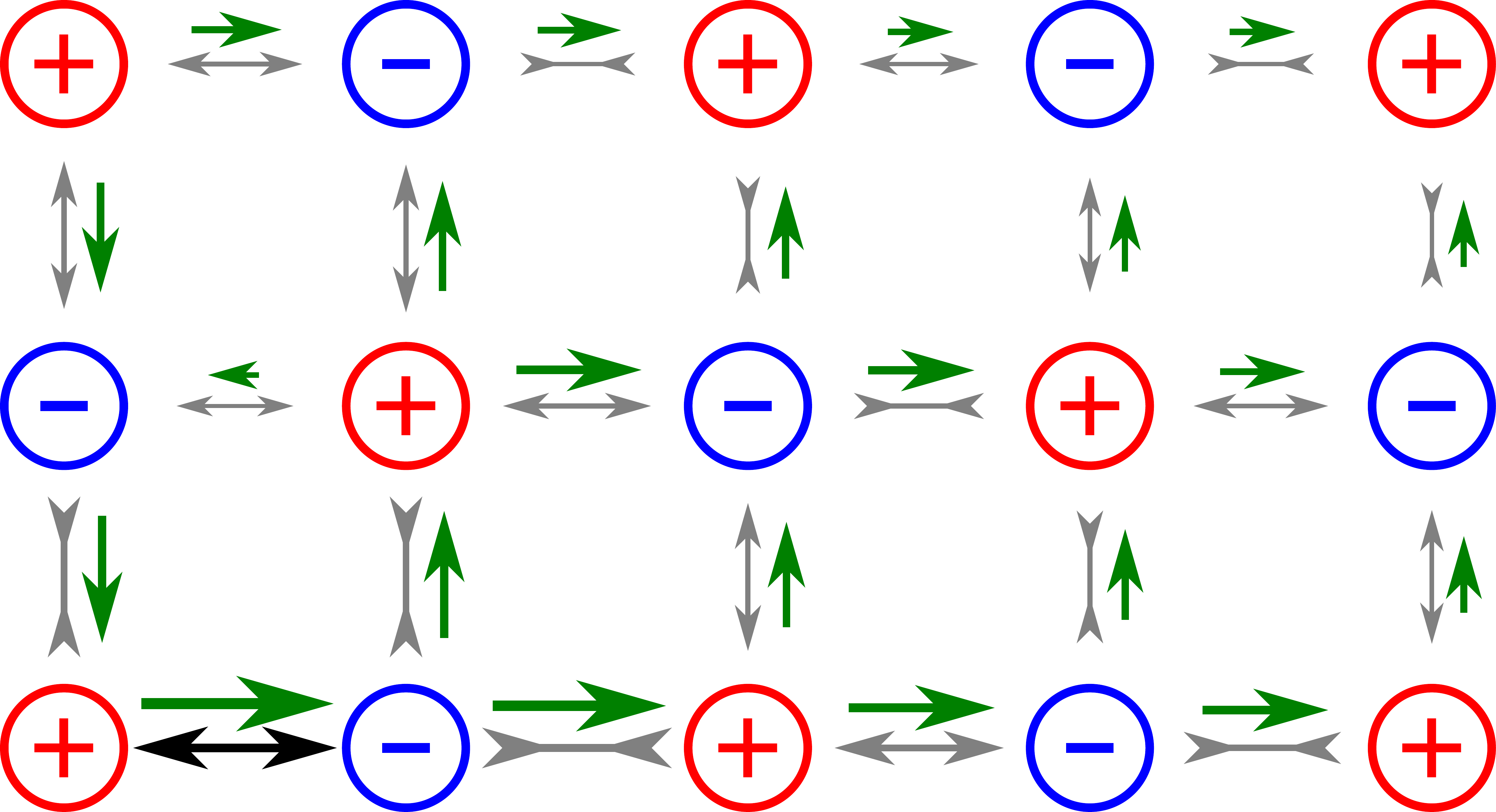}
	\caption{Representative picture of the long and short bond arrangement. Red circles represent Pb atoms, blue circles Te atoms. The reference bond (black) is assumed to be longer than average. Gray standard arrows (pointing outwards) mean a positive correlation with the reference bond; if the first bond is longer, the second tends also to be longer. Gray arrows pointing inwards mean a negative correlation with the reference bond. The size and thickness of the arrows are indicative of the strength of the correlations. Green arrows indicate the size and direction of the resulting dipole moments.}
	\label{fig:dips-picture}
\end{figure}%
The horizontal $[100]$ line containing the reference bond shows a clear alternation of short and long bonds (i.e.\ negative correlations), resulting in a \textit{local} \quotemarks{ferroelectric-like} arrangement of electric dipoles. 
The neighboring, parallel $[100]$ lines show the same pattern with a \quotemarks{ferroelectric} coupling to the reference $[100]$ line. 
The only exception is the $B^4_3$ bond which has a weak tendency to a positive correlation -- strain and dipolar interaction are strong enough to counterbalance the ferroelectric coupling that would favor a negative correlation. 
On the other hand, the vertical $[010]$ line also presents a ferroelectric-like arrangement of bonds, but with the formation of a \quotemarks{head-to-head domain wall} at the reference $[100]$ line. 

Note that the formation of ferroelectrically coupled ferroelectric-like $[100]$ lines is not expected from the long-range dipole-dipole interaction, which would favor an antiferroelectric coupling of ferroelectric $[100]$ lines. 
The origin of such a coupling may be found in short-range interactions depending on the chemistry of the environment, in this case the stereochemical activity of the lone pair. 
A similar competition was noted for the bulk ferroelectric behavior in perovskite BaTiO$_3$ by \citet{Nishimatsu2010} with the construction of an effective Hamiltonian\cite{Zhong1995}. The authors showed that when only the dipole-dipole interaction is considered the most unstable phonon mode corresponds to the $M$ point -- the most stable structure consists of an antiferroelectric cell-doubling state. Only when the short-range interactions are also included is the most unstable mode found at $\Gamma$ producing the actual ferroelectric state. 

At first sight our findings in this section could seem in contradiction with our earlier discussion that the average structure remains high-symmetry rocksalt. 
However, the bond lengths averaged over all higher-order correlations are unchanged from those of the reference state, with the most probable bond length still found at $a/2$ (see Figs.~\ref{fig:app-corrs4-neut}-\ref{fig:corrs4} in the Appendix). This hints at a reconciliation between the conflicting descriptions -- anomalous anharmonicity or off-centering -- which we will explore further in the next section.

\subsection{Local off-centering}
\label{sec:offc}

\begin{figure}[hptb]
	\includegraphics[width=.95\columnwidth]{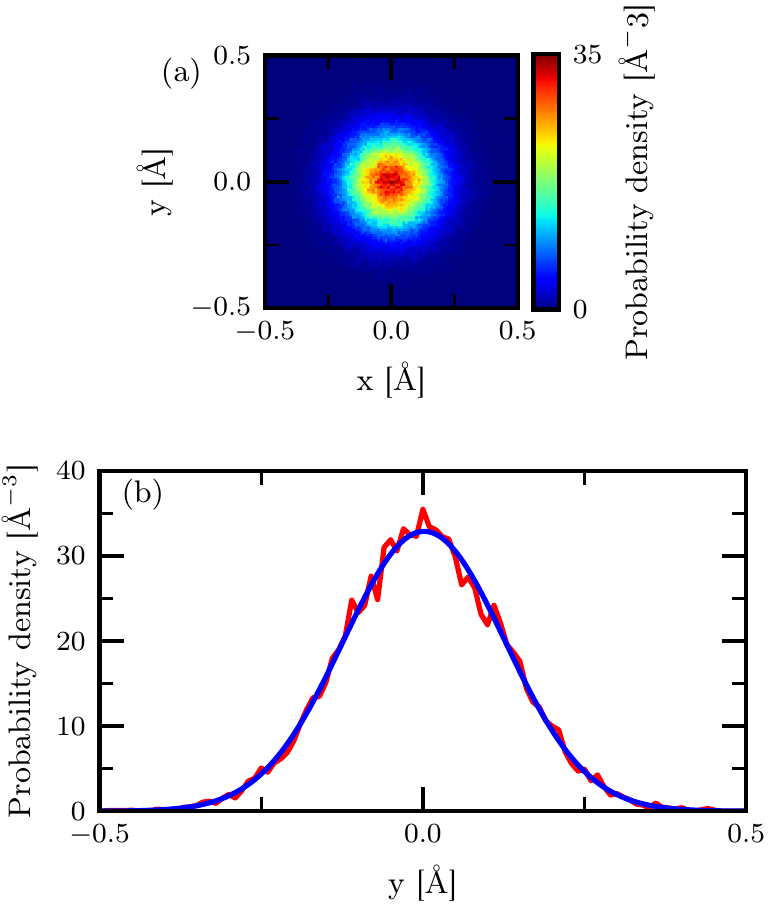}
	\caption{
    Probability density distribution for local lead off-centering from our ab-initio MD simulations. (a) shows a cut in the $x-y$ plane, while (b) a cut for $x=z=0$. The red line represents the actual data, the blue line a Gaussian fit.
	}
	\label{fig:lead-offc}
\end{figure}

Next we discuss the implications of these correlated dipoles in the context of the proposed off-centering. In particular, we show that the existence of correlated local dipoles is consistent with an average local Pb position which is centered in the Te coordination polyhedron. 
Figure \ref{fig:lead-offc} presents the deviations of the Pb atoms from the center of gravity of the PbTe$_6$ octahedra. 
Since it is also a many-body problem this information is not directly accessible from the diffuse scattering or the 3D-$\Delta$PDF data. Therefore we show only results from our MD simulations, which we justify by the good agreement in the local structure between MD and 3D-$\Delta$PDF. 
Clearly, the distribution is centered on the origin and a Gaussian profile can not be excluded. 
Here, only the $x-y$ plane is shown but the same conclusions are obtained also for all inequivalent $[100]$, $[110]$, and $[111]$ directions. 
The same picture is obtained when considering the displacement of Te atoms with respect to the TePb$_6$ octahedra. 
So, we see that \textit{on average} both Pb and Te atoms are not off-center but sit in the center of their coordination polyhedra, consistent with our previous finding of the most probable bond being at $a/2$ in the higher-order correlations (Figs.~\ref{fig:app-corrs4-neut}-\ref{fig:corrs4}). 

We next link this result with our previous discussion about correlated dipoles. 
In a specific unit cell, as we have just seen, the probability distribution for distortions is Gaussian without any preferential direction. 
As soon as one unit cell has chosen a distortion direction, however, the neighboring unit cells are no longer free to choose their distortion directions, since the pair and higher-order correlations impose a preferred orientation. The result is the development of regions of local polarity composed of correlated dipoles along the $\langle 100 \rangle$ directions.
These distorted regions are randomly distributed in the crystal, however, such that averaging over them results in an (undistorted) rocksalt structure. 
From the decay of the higher-order correlations' strength we can roughly estimate the extent of these correlated dipoles regions to be around \mbox{$30$~\AA}. 
These regions may resemble the dynamically fluctuating polar nanodomains proposed in relaxor ferroelectrics, as in the analysis of the diffuse scattering by Bosak \textit{et al.}~\cite{Bosak2012}.

We emphasize the distinction from a static local off-centering, in which the interatomic potential would not have a single minimum, but would show other (meta-)stable states. This could result in an alternation of long and short bonds, with the formation of one shoulder on each side of the PDF peaks, or even of multi-valued peaks. 
Moreover, the distribution of the deviations with respect to the coordination polyhedra shown in Fig.~\ref{fig:lead-offc} would not be Gaussian and may show multiple peaks. 

We propose that our finding of correlated local dipoles combined with centered average Pb environments automatically resolves some of the controversy in the literature, which has been caused by differing interpretations of the meaning of ``off-centering'', such as the root mean square\cite{Kim2012} or absolute\cite{Zhang2011,Chen2014,Shiga2014,Li2014b} displacement from the rocksalt positions. 
Here we show that the phenomenon called emphanisis is associated with the formation of correlated local dipoles which can both vary in orientation throughout the structure and fluctuate in time. As a result the positions of the ions are centered on average. The effect might be better referred to as \textit{correlated local dipole formation} to avoid confusion in future works.

\subsection{Radial pair distribution function analysis}

\begin{figure}[hptb]
	\includegraphics[width=.95\columnwidth]{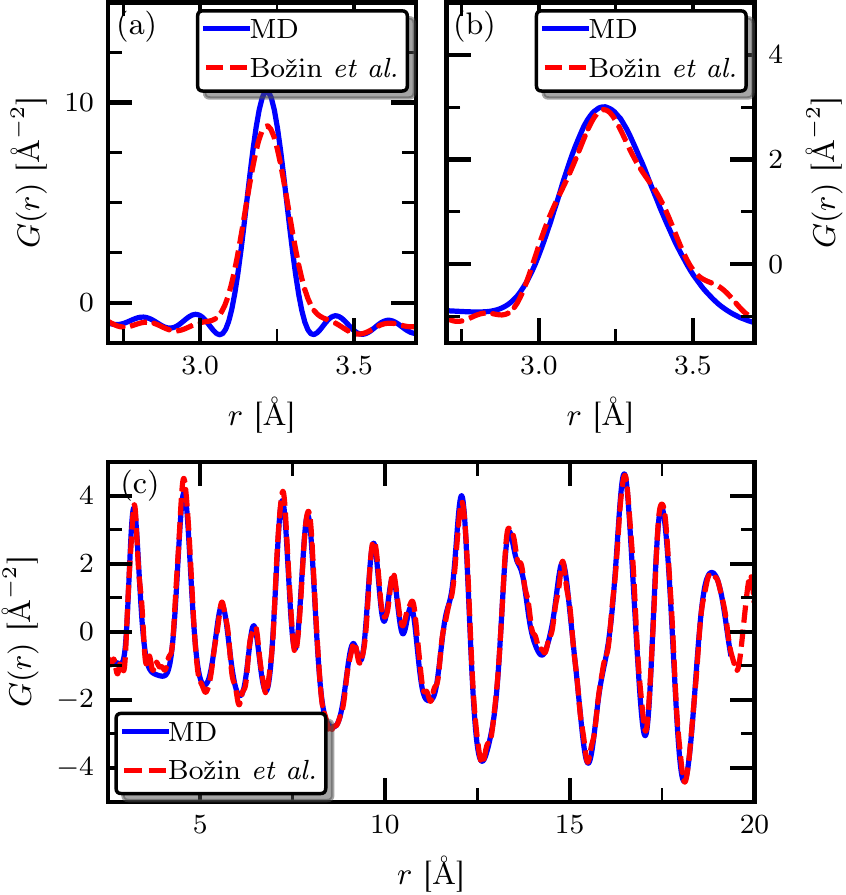}
	\caption{Comparison between the PDFs obtained from our MD simulations (blue lines) and the original data from  \citet{bozin2010} (red dashed lines) with the temperature correction suggested in Ref.~\onlinecite{Knight2014}; (a) $T=50$ K, (b) $T=450$ K, and (c) $T=300$ K (the larger supercell was used). The MD data were multiplied by a scaling factor of $0.89$. 
	}
	\label{fig:1peak-cf}
\end{figure}

\begin{figure}[hptb]
	\includegraphics[width=.95\columnwidth]{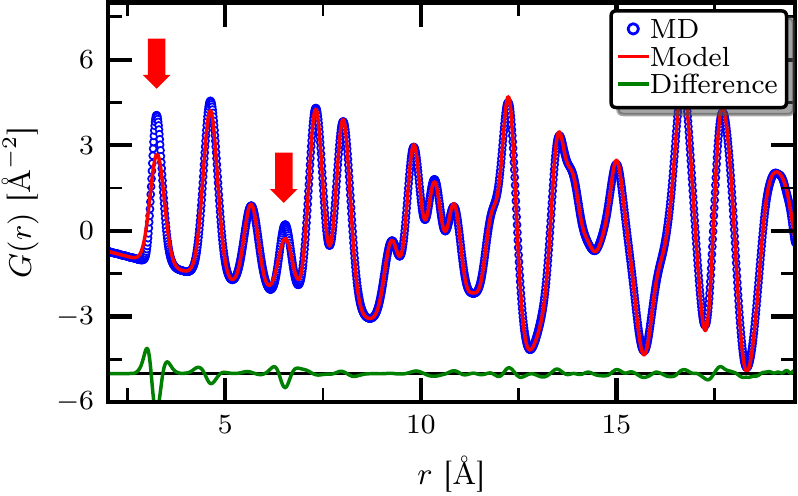}
	\caption{Comparison of the refinement of the MD PDF at $300$~K with that of a cubic $Fm\bar3m$ model without correlated motion taken into account.
	}
	\label{fig:pdf-refinement}
\end{figure}

Finally we show that our finding of local correlated dipoles is largely consistent with the peculiar features of the original pair distribution function (PDF) of \citet{bozin2010} with the exception that the shoulders in the peaks are not reproduced by our MD simulations. 
To do this, we first compare the radial PDF $G(r)$ extracted from our MD simulations with the original data. We computed the PDF from the MD as a histogram of the interatomic distances and by taking into account the neutron scattering lengths ($9.405$ fm for lead and $5.80$ fm for Te\cite{Sears1992}), with the experimental termination error from the finite $Q$ range of the Fourier transform ($Q_\text{max}$ set to $35$ \AA$^{-1}$), and scale by a factor of 0.89 to take into account the arbitrary scaling introduced by some of the applied data corrections~\cite{Juhas2013}. 
Figure \ref{fig:1peak-cf} compares the MD and measured PDFs at three different temperatures, $50$~K (a), $450$~K (b) and $300$~K (c). 
As can be seen, at the lowest temperature shown, $50$~K (panel(a)), our MD calculations underestimate slightly the width of the nearest-neighbor peak -- this may be ascribed to the underestimation of the ADPs originating, at this low temperature, from the lack of zero-point motion in the MD simulations. 
On the other hand, at the other two temperatures shown the agreement is remarkably good. 
Only the shoulder on the high-$r$ side of the peak at $450$~K (panel (b)) is not captured. 
(The two small shoulders appearing at approx.\ $2.8$ and $3.6$~\AA\ in the experimental data are probably artifacts; their position is consistent with the periodicity of the wiggles from the finite $Q$ range of the Fourier transform.)  In particular, Fig.~\ref{fig:1peak-cf}(c) shows that our MD simulations at 300~K capture not only the correlated motion of the nearest-neighbor peak at $\sim3.2$ \AA, but also the intriguing overly sharp lattice-repeat-distance peak of the original data. 
In Fig.~\ref{fig:pdf-refinement} we compare our 300K MD data with a model that neglects correlated motion. We find the largest differences for the first and fourth peaks (marked by the red arrows in the figure), which are sharper (their calculated width, not shown, is smaller) than the second, third, fifth, and so on. 
The sharper peaks correspond to interatomic distances between atoms lying along the $\langle 100 \rangle$ directions, and their sharpness is a result of the pair correlations being positive, and strongest along the $\langle 100 \rangle$ direction. 

\begin{figure}[hptb]
	\includegraphics[width=.95\columnwidth]{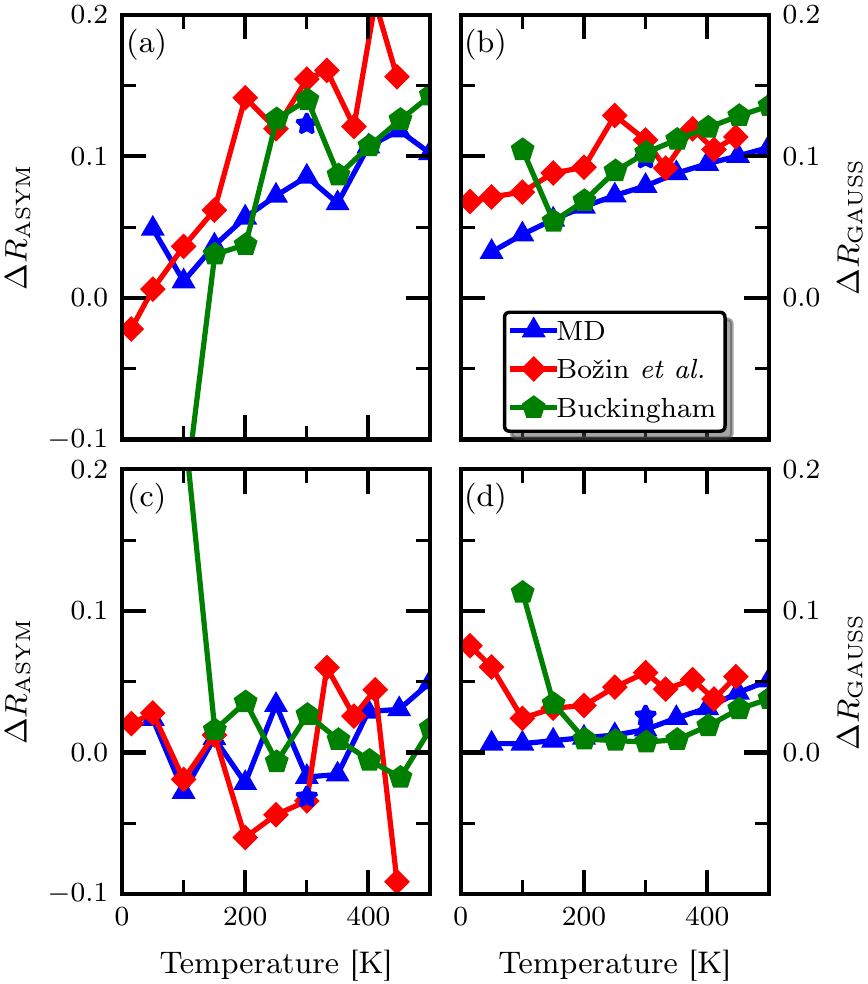}
	\caption{Asymmetry parameters of the first two peaks of the RDF as a function of temperature. 
    The top panels (a)-(b) show the asymmetry parameters for the nearest-neighbor peak, while the bottom panels (c)-(d) for the next nearest neighbor peak. The left panels (a) and (c) present the asymmetry parameter $\Delta R_\text{ASYM}$, while the right panels (b) and (d) $\Delta R_\text{GAUSS}$. 
    Blue symbols are obtained from our \textit{ab-initio} MD simulations (blue stars correspond to the larger supercell), red symbols are from the original data of \citet{bozin2010}, and green symbols are from the Buckingham potential.}
	\label{fig:asym}
\end{figure}

While visual inspection suggests that the MD peaks are less asymmetric than the data, our quantitative analysis shows that the asymmetry is actually captured quite well, in particular the trend towards more asymmetric peaks with increasing temperature. 
For the computation of the asymmetry we first transform the PDF $G(r)$ to the radial distribution function (RDF) $R(r)$, such that its peaks' shapes describe the underlying pair-probability distribution (for a Gaussian probability distribution the peaks are symmetric) and the baseline lies at zero. We define asymmetry in two different ways, the first being %
\begin{equation*}
\label{eq:asym-1}
\Delta R_\text{ASYM}=\frac{\int_0^{\delta r}\left[ R(r_0+r) - R(r_0-r)\right]\,\mathrm dr}{\int_0^{\delta r} \left[ R(r_0+r) + R(r_0-r)\right]\,\mathrm dr}\,,
\end{equation*}%
the second defining a \quotemarks{relative error} with a Gaussian function, %
{\small%
\begin{equation*}
\label{eq:asym-2}
\Delta R_\text{GAUSS} = \frac{\int_{r_0-\delta r}^{r_0+\delta r} \left| R(r) -  N_c\cdot\mathrm{e}^{-(r - \mu)^2/2m_2}/\sqrt{2\pi m_2}\right|\,\mathrm dr}{\int_{r_0-\delta r}^{r_0+\delta r} \mathrm{e}^{-(r - \mu)^2/2m_2}/\sqrt{2\pi m_2}\,\mathrm dr }\,,
\end{equation*}%
}%
where $N_c=\int_{r_0-\delta r}^{r_0+\delta r}R(r)\,\mathrm dr$ is the coordination (area of the peak), and $m_2$ is the variance of the peak. 
If the peak is not only asymmetric but also displays shoulders, the Gaussian parameter, $\Delta R_\text{GAUSS}$ should be a better indicator than others relying on a difference between high- and low-$r$ sides. 
Figure \ref{fig:asym} shows the temperature evolution of the two asymmetry parameters calculated from our \textit{ab-initio} MD simulations (blue), from the original data by \citet{bozin2010} (red), and from the Buckingham potential (green).  
Both asymmetry parameters deliver a constantly increasing asymmetry with temperature for the nearest neighbor peak in good agreement with the experimental data by \citet{bozin2010}. The next nearest neighbor peak is more symmetric, but shows increasing deviations from a Gaussian profile too. 
Note that using the skewness as a measure of the asymmetry leads to the same conclusions. 

\begin{figure}[hptb]
	\includegraphics[width=.95\columnwidth]{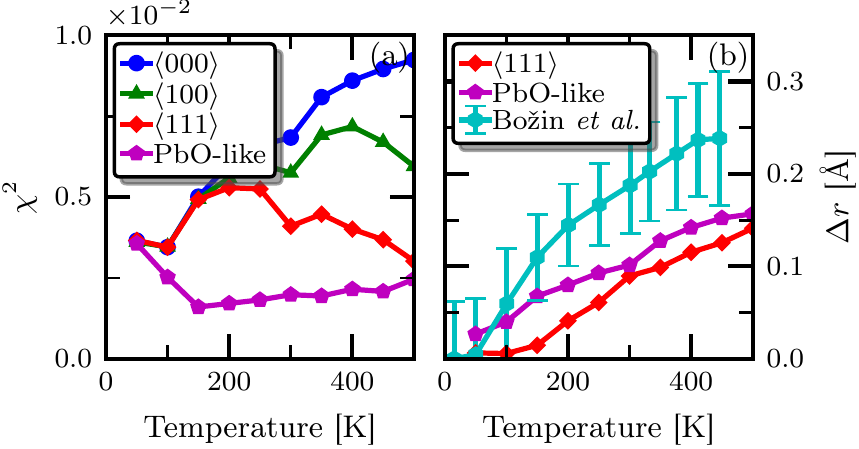}
	\caption{Comparison of different structural models fit to our MD simulations. (a) Quality of the fit through the $\chi^2$ goodness parameter; (b) \quotemarks{off-centering} displacement, $\Delta r$, extracted for the two most probable (split) models compared to the data extracted from the study by \citet{bozin2010} with the temperature correction proposed in Ref.~\onlinecite{Knight2014}.}
	\label{fig:pdf-refinement-offc}
\end{figure}

Next, we subject the PDFs obtained from our MD simulations to the same small-box fitting protocols as in the original PDF study \cite{bozin2010}. We used the PDFgui package\cite{Farrow2007} and set all experimental setup related parameters (scale factor, $Q$-space resolution and related damping in $r$-space) to ideal values. Various models were refined over a narrow $r$-range up to $6$~\AA, some with lead displacements allowed and some not: the cubic $Fm\bar3m$ rocksalt ($\braket{000}$); $\braket{100}$ and $\braket{111}$ models, where the lead sublattice is allowed to shift with respect to the Te sublattice in the specified direction; a $\braket{100}$ PbO-like model, where a tetragonal cell is used and lead is allowed to displace in a pattern similar to that in PbO. In Fig.~\ref{fig:pdf-refinement-offc}(a)) we show the quality of the fits, measured by the $\chi^2$ parameter, as a function of temperature for the various models. We see that the cubic model does progressively worse with increasing temperature, while the best fit is produced by the $\braket{100}$ PbO-like model in particular at higher temperatures, consistent with the original experimental observation \cite{bozin2010}. 
Note that the $\braket{111}$ model also gives reasonably good fits. 
Next we extract an \quotemarks{off-centering} displacement from the two best models and show our results in Fig.~\ref{fig:pdf-refinement-offc}(b). 
The resulting values are in good agreement with the original reported values in Ref.~\onlinecite{bozin2010} (light blue symbols).  

While this finding of off-centering might seem to contradict our finding above that the Pb ions are at the center of their Te coordination polyhedra, in fact it is a consequence of the interpretation and use of split models in current real-space refinement utilities. 
In fact, the small-box modeling software, PDFgui, is not designed to handle anharmonic effects, and instead uses Gaussian distribution shapes. 
Instead of accounting for anharmonic effects directly, these effects are handled indirectly via discrete bond-length distributions and partial occupancies (split models). 
Consider the situation shown in Fig.~\ref{fig:anh-sketch}: 
\begin{figure}[hptb]
	\includegraphics[width=.95\columnwidth]{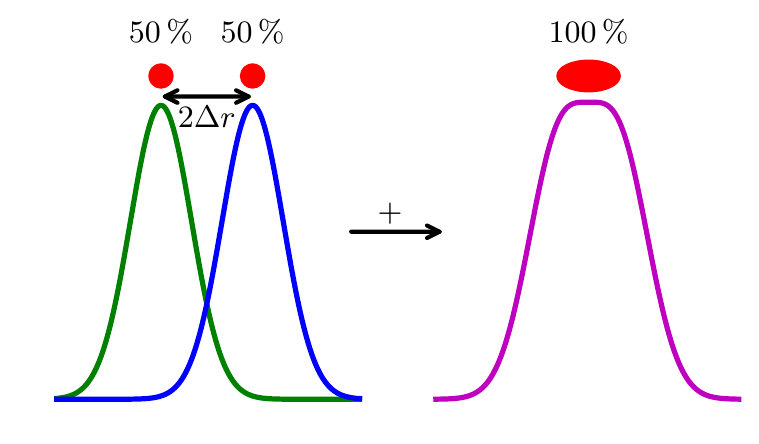}
	\caption{Representation of how anharmonicity can be introduced in real-space refinement utilities through split models. 
	}
	\label{fig:anh-sketch}
\end{figure}
on the left a split model is considered, with each split position ($50\,\%$ occupancy) described by a Gaussian and separated by the \quotemarks{off-centering} $2\Delta r$. However, the sum of the two Gaussians  (right) results in a pseudo-Gaussian distribution that mimics an anharmonic potential. 
This means that if the final distribution of a split model is not multivalued, anharmonicity and off-centering can not be distinguished. 
It would be therefore desirable to upgrade real-space refinement procedures with anharmonic features to avoid confusion in the description of strongly anharmonic systems such as the group IV chalcogenides.

\subsection{Origin of the peak asymmetry}
\label{sec:origin-asymmetry}

Finally, we discuss the origin of the unusual behavior and conflicting literature reports in light of our new measurements and calculations. 
First we address the relationship between the asymmetry of the peaks and the anharmonicity of the PbTe potential. 
In the previous studies that did not find any off-centering \cite{Zhang2011,Chen2014,Shiga2014,Li2014,Li2014b} , the well-known anharmonicity of PbTe was considered to be the sole cause of asymmetry. 
Indeed, an anharmonic potential, such as the Buckingham potential used above, can explain alone (part of) the asymmetry. 
Figure~\ref{fig:asym} shows that also the Buckingham potential produces an increased asymmetry with good agreement with the data from Ref.~\onlinecite{bozin2010}.  
On the other hand, anharmonicity alone is not enough to explain the peculiar decay of the pair correlations presented in Fig.~\ref{fig:corr-dft-1}, since we showed that the Buckingham potential does not lead to steps in the pair correlations  (see Fig.~\ref{fig:buck_corrs}). 

\begin{figure}[hptb]%
	\includegraphics[width=.95\columnwidth]{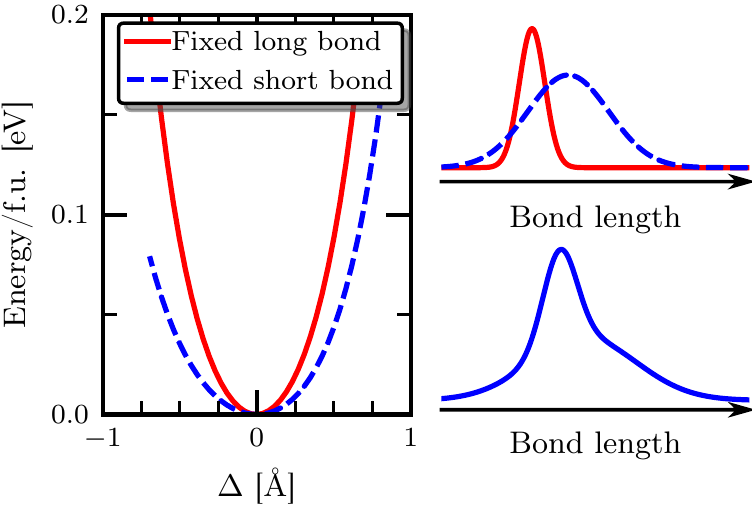}
	\caption{ Left panel, energy change when either the long ($3.5$ \AA) or the short ($2.94$ \AA) bond are kept fixed, and the other is varied. $\Delta$ represents the change in bond length with respect to the equilibrium bond length (the bond corresponding to the minimum energy). The calculation was done using VASP by building a $3\times1\times1$ cubic supercell and displacing two atoms to produce the sought for bond arrangement. Right panel, top, sketch of the probability distributions of the short (red) and long (dashed blue) bonds, bottom, sum of the two distributions. 
	}
	\label{fig:asym-bonds}
\end{figure}

Next, we show that the correlated dipoles can further amplify the peak asymmetry. Figure \ref{fig:asym-bonds} shows results of calculations using VASP for a cubic supercell containing one long and one short bond. 
The length of these two bonds was chosen to differ by about $10\,\%$ from the equilibrium bond length, giving values of $3.5$ \AA\ and $2.94$ \AA. The large weight in the  probability densities of Fig.~\ref{fig:corrs3}(d)-(e) corresponding to these bond lengths shows that they are a reasonable choice. 
One of the bonds was changed, while the other was kept fixed, and the energy was calculated. 
One can see that when the short bond is varied (fixed long bond) the energy curve is steeper than when the long bond is changed (fixed short bond), because of the asymmetry of the crystal field. 
This produces a narrower distribution for short bonds than long ones, resulting in an overall asymmetric distribution of bonds (see right panel in Fig.~\ref{fig:asym-bonds}). 
Thus, the alternation of short and long bonds provides another microscopic contribution to peak asymmetry. 

We note that the previously reported shoulder on the first PDF peak, which was interpreted as a local off-centering of the Pb-Te bonds, is not reproduced either in our molecular dynamics simulation or diffuse scattering (nor elsewhere in the literature). Therefore, any additional physics associated with the existence of this peak are not captured by our analysis.  Possible reasons for its absence in our molecular dynamics simulations are an insufficiently large supercell, or the absence of spin-orbit coupling in the calculations. 
We note, however, that the existence of such a shoulder is not essential for emphanitic behavior, which we find in our MD simulations and which has been reported for other materials such as CsSnBr$_3$ \cite{Fabini2016}, where shoulders are lacking.

\section{Relevance for thermoelectricity}
\label{sec:thermoelectricity}

In this last section, we discuss the relevance of our findings for the well-known thermoelectric performance of PbTe  \cite{Wood1988,Heremans2008,Androulakis2010,Zhao2013,Goncalves2014,Zhao2014a,Han2016,Tan2016a}. Clearly the fluctuating correlated local dipoles will contribute to phonon scattering; here we show that they should also influence the electronic behavior. 
It is well-known that lone-pair expression is beneficial for thermoelectric performance \cite{Nielsen2013a,Zeier2016}, since it tends to open electronic  band gaps and suppress bipolar conduction of carriers. 
The electronic structure of PbTe indicates a further mechanism to increase its performance, that is the presence of a secondary pocket along the $\Sigma$ line with an energy slightly lower than those of the $L$ pocket, where the band gap can be found \cite{Giraldo-Gallo2016}. 
By reducing the band offset between $L$ and $\Sigma$ pockets the valley degeneracy can be increased ($L$ pockets have a valley degeneracy of $4$, while that of $\Sigma$ pockets is $12$), and consequently also the power factor\cite{Androulakis2010,pei2011,pei2012,He2013,Zhao2013a,Zhao2013,zhu2014,Zeier2016}. 
Experimentally, in pristine PbTe this decrease in the offset happens with increasing temperature, with a proposed convergence at about $700-800$~K\cite{Gibbs2013,Zhao2017}. 
The band offset can also be controlled by K or Na co-doping such that the thermoelectric power factor is enhanced through the tuning of the interaction between the $L$ and $\Sigma$ bands \cite{Androulakis2010}. 

\begin{figure}[hptb]
	\includegraphics[width=.95\columnwidth]{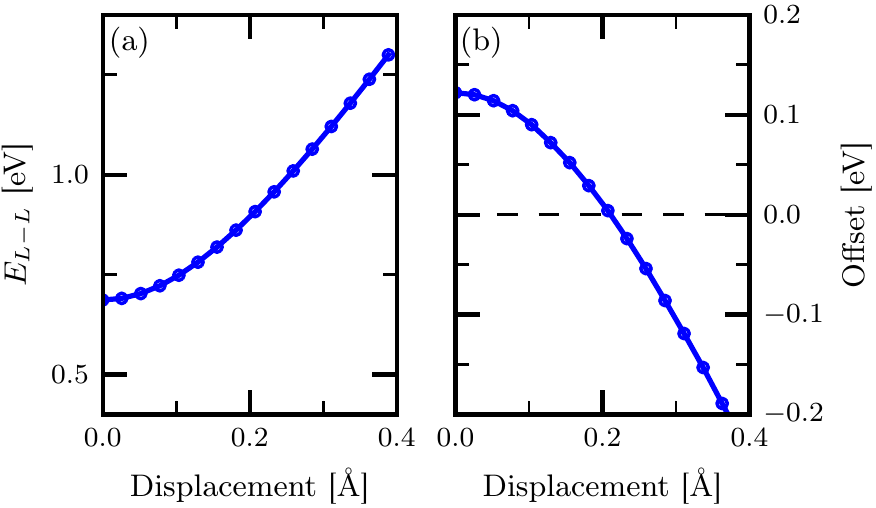}
	\caption{Electronic properties as a function of the ferroelectric distortion along $\langle 100\rangle$. (a) Direct band gap at $L$ and (b) offset between $L$ and $\Sigma$ pockets. 
	}
	\label{fig:band-gap}
\end{figure}

In an attempt to link the correlated dipoles to the high thermoelectric performance of PbTe we show in Fig.~\ref{fig:band-gap} the band gap and band offset between $L$ and $\Sigma$ pockets as a function of a polar distortion along $\langle 100\rangle$. 
Here we use a structure with the extreme situation of a ferroelectric distortion along $\langle 100\rangle$, with aligned static stereochemically active lone pairs\cite{Nielsen2013a}.  
As expected, we find that lone-pair expression is accompanied by an increase of the band gap (panel (a)). 
More importantly, the band offset (panel (b)) decreases with increasing distortions, so that the band gap becomes indirect for displacements above about $0.2$ \AA. 
While the cooperative ferroelectric distortion analyzed here of course differs from the actual correlated local dipole behavior, we expect that the trends will be similar, with the amplitude of distortion in our calculations playing the role of temperature in the experiments.  
Thus, the correlated dipole regions likely contribute to the high thermoelectric performance of PbTe by affecting both the electronic and phononic components.

\section{Summary}
\label{sec:summary}

In summary, we have performed a combined x-ray scattering and \textit{ab initio} molecular dynamics study of the lattice behavior of PbTe, and analyzed our results in the context of earlier PDF studies that suggested a local Pb-Te off-centering. The presented 3D-$\Delta$PDF analysis provides a detailed quantitative picture of the atomic pair correlations up to distances of about 60 \AA. This allows a comprehensive insight into the local structure of PbTe, and also demonstrates the power of the 3D-$\Delta$PDF method for analyzing the local structure of disordered crystals. 

We find a close consistency between our Bragg and diffuse x-ray scattering measurements and our molecular dynamics simulations. At the bulk level, both our Bragg scattering measurements and our calculations yield non-Gaussian peaks, indicative of significant anharmonic behavior. 
At the local level, both our diffuse scattering and MD simulations find a distinctive and unusual stepwise decay with distance in the pair correlation functions. This excellent match allows us to further analyze our molecular dynamics results to extract detailed information about the atomic positions and dynamics -- including higher-order correlation functions -- that can not be easily obtained from experiment. 

Our main finding is the unusual correlated local dipole formation, exemplified by Fig.~\ref{fig:corr-dft-1}, in which a displacement of an ion along a cubic axis causes correlated displacements in the atoms in neighboring cells that do not show the usual smooth decay with distance. Instead, the correlations indicate a tendency for anion-cation dimerization, and the resulting formation of local dipoles, in the direction of displacement. This behavior can be interpreted as a superposition of an acoustic phonon mode, which alone would displace all atoms in the same direction and would decay smoothly with distance, with an optical phonon mode, which causes opposite atomic displacements and therefore is responsible for the step-like behavior. Thermal activation of an optical phonon mode is clearly required for such an effect, and is possible in PbTe because of the soft transverse optical phonon mode associated with the proximity to the ferroelectric phase transition. Since the essential ingredient for the observed local structure is a soft optic phonon mode with a strong coupling with acoustic modes, similar behavior should occur, in principle, close to any ferroelectric phase transition. In this respect, it would be interesting to search for such correlated local dipole behavior above the transition temperature in ferroelectric phase transitions that have been previously regarded as displacive. The finding using EXAFS spectroscopy that the local distortions associated with the ferroelectric state in GeTe persist essentially unchanged on heating to the paraelectric phase, might indeed be an indicator of emphanisis~\cite{Fons2010}.

Importantly, this correlated dipole formation does not translate into an \textit{average} off-centering of the ions from the centers of their coordination polyhedra. Analysis of our simulations indicates that the most probable ionic position, averaged over time or space, is at the center of the polyhedron, with a smooth decrease in probability away from that point. This is consistent with different regions of the material having different orientations and amplitudes of correlated dipole formation, and with a local but not global symmetry breaking of the cubic symmetry. The question of \quotemarks{off-centered or not} that has recurred throughout the literature is therefore perhaps not the right question to ask in this case. 

Finally, we suggest that the correlated dipoles and the associated distortions along $\langle100\rangle$ are beneficial for the high thermoelectric performance of PbTe. 
First, the coupling between acoustic and optic phonons (as was already shown in Ref.~\onlinecite{Delaire2011}) and the  regions with different orientations of the correlated dipoles in the crystal may increase the phonon scattering, consequently decreasing the thermal conductivity. 
Second, the distortions along $\langle100\rangle$ may explain two electronic features that are thought to be necessary to explain the thermoelectric performance, the increase of the band gap with temperature and, at the same time, the band convergence of the $L$ and $\Sigma$ pockets.

\begin{acknowledgments}

BS, MF and NAS acknowledge support from ETH Z\"urich, the ERC Advanced Grant program (No. 291151), and the Swiss National Supercomputing Centre (CSCS) under project IDs s307, s624 and p504. We thank Joost VandeVondele for helpful discussions, in particular regarding the use of the CP2K code. 
TW thanks the staff of the X06SA beamline at the Swiss Light Source, Villigen, Switzerland for giving access to the beamline and for helping with the experiments. 
Work at Brookhaven is supported by the U.S.~DOE under contract No.~DESC$00112704$. 

\end{acknowledgments}
	
\bibliography{Bibliography} 

\vspace{2cm}

\appendix
\section*{Appendix}\label{sec:appendix}
\setcounter{figure}{0} 
\setcounter{table}{0} 
\makeatletter 
\renewcommand{\thefigure}{A\@arabic\c@figure}
\renewcommand{\thetable}{A\@Roman\c@table}
\makeatother

\section{Methods}
\label{sec:methods}

\subsection{Sample Preparation}
Single crystals of PbTe were prepared by mixing stoichiometric amounts of Pb (Rotometals, at 99.9 \%  purity) and Te (Plasmaterials, at 99.999 \% purity) in an evacuated fused silica ampoule. The total amount of PbTe was 15.346 g and the length of the ampoule was approximately 18 cm. The ampoule was placed in a box furnace, heated to 1050 $^\circ$C over 16 h, and held at that temperature for 36 h. The sample was then cooled to room temperature in 24 h. Small single crystals ($10-100$ $\mu$m) were formed on the top part of the ampoule. Several crystals were screened on a STOE IPDS 2T diffractometer for quality evaluation.

\subsection{X-ray diffraction}
The single crystal used for the X-ray experiments was an irregularly shaped fragment with an average diameter of about $42$~$\mu$m. The  experiment was done at the X06SA beamline at SLS, Villigen, Switzerland, which was equipped with a single axis goniometer and a PILATUS 6M detector. The synchrotron storage ring was operated in top-fill mode to deliver constant primary intensities. Full data sets (3600 frames, $0.1\degree$ rotation and 0.1~s exposure time per frame, wavelength 0.7085~\AA, crystal - detector distance 185 mm) were recorded in the sequence room temperature, 250 K, 200 K, 150 K and 125 K. The experimental setup did not allow access to temperatures above room temperature. The primary beam was normal to the rotation axis and to the detector plane. The detectors' energy threshold for accepting X-rays photons was set to 16 keV (energy of primary beam $17.5$ keV) to suppress fluorescence scattering as much as possible. In addition, 100 frames were collected under the same conditions as in the diffuse scattering measurements, but without sample and sample holder. These frames were averaged and taken as a model for background scattering.

A Bragg data set collected with a strongly attenuated beam did not deliver satisfactory results, as the internal R-value was well above $0.2$ (the internal R-value is defined as $R_{int}=\frac{\sum |I_{obs}-\langle I_{obs}\rangle |}{\sum |I_{obs}|}$, where the summations run over all reflections for which more than one symmetry equivalent reflection is averaged, $I_{obs}$ is the corrected intensity of a reflection and $\langle I_{obs}\rangle$ is the corresponding symmetry-averaged intensity) even for the triclinic Laue group. Reasons may be strong multiple scattering effects and/or unresolved saturation problems of the detector. We therefore repeated the Bragg data collation  using our in-house Xcalibur Oxford Diffraction diffractometer (Mo K$_\alpha$  radiation, graphite monochromator, sealed tube, Onyx CCD detector, $\theta_{max}=45.3^{\circ}$, $-12 \le h,k,l \le 12$ same crystal and temperatures as used in the synchrotron experiments).

\subsection{Diffuse scattering data reduction}
Reciprocal space reconstructions were done with the program XCAVATE\cite{Estermann1998}. Reconstructions were mapped onto a $360\times360\times360$ voxel volume covering the reciprocal space within the limits $-9 \le h,k,l \le 9$. Since the half-widths of the Bragg reflections were significantly smaller than the pixel size of the reconstructions, it was assumed that the experimental reciprocal space resolution function can be ignored to a good approximation. In contrast to the Bragg data, the diffuse scattering was corrected using a spherical absorption model for sake of simplicity. It is not expected, however, that this simplification will introduce significant systematic errors. For comparison, Bragg data corrected by spherical absorption correction resulted in slightly higher $R_{int}$ values (0.11 vs. 0.078 using the analytical approach, see below), but the refined structure parameter values from those data changed only by a few percent compared to the results obtained with analytical absorption correction. It is not expected that the choice of the absorption correction method would have a large impact on the results extracted from the diffuse data. Finally, the data were averaged following the Laue symmetry $m\bar{3}m$. Individual voxels were rejected as outliers according to the procedure described in Blessing\cite{Blessing1997}. A voxel was rejected if its difference to median value of symmetry equivalent voxels was more than about six times the median difference (for details see Ref.~\onlinecite{Blessing1997}, Eqs.~16, 17 with $c_1, c_2, c_4 = 0$,  $c_3 = 6$). This procedure turned out to be very helpful in eliminating most of the parasitic scattering that is not compatible with the Laue symmetry of the single crystal diffraction pattern, such as scattering from small grains attached to the surface or secondary air scattering from strong reflections. Finally, the background was subtracted from the diffuse data. It was expected that the empirical experimental background obtained as described above would show slightly smaller background intensities compared to the background seen with the crystal, because scattering from the sample holder and from glue were not included in the background measurements. To our surprise we found, however, that the empirical background determination showed slightly larger intensities compared to those observed in the  diffuse scattering measurements. This observation may be explained by the fact that the primary beam gets attenuated by the crystal and therefore the air scattering coming from the pathway between crystal and beam stop is reduced. To partly compensate this effect we multiplied the empirical background by a factor of 0.9 to avoid larger regions of negative intensity after correction.

\vspace{-0.3cm}

\subsection{Bragg data absorption correction}
The high absorption coefficient of $\mu=75.5$ mm$^{-1}$ required careful absorption correction. Various strategies provided by the program CrysAlisPro\cite{Crysalis} were tried, namely analytical absorption correction based on a graphical estimate of the crystal's morphology and its orientation relative to the diffractometer coordinate system, spherical absorption correction, empirical absorption correction and combinations of analytical/empirical, spherical/empirical and analytical/optimised crystal shapes approaches. The best internal R-value ($R_{int}$ = 0.064) was obtained from the analytical/optimised crystal shapes method, however, the optimised crystal shape did not well match the morphology of the sample and the results were therefore rejected to avoid overcorrection. The second best results were obtained from the analytical approach and from the combination analytical/empirical absorption correction ($R_{int}$ = 0.078 in both cases). Since the latter did not deliver better data, the results obtained from the analytical correction method were used in our refinements.

\vspace{-0.28cm}

\subsection{Computational details}
\label{sec:computational-methods}

Our ab-initio molecular dynamics simulations were performed using the CP2K code \cite{cp2k} with the hybrid Gaussian and plane wave (GPW) scheme \cite{lippert1997,VandeVondele2005}. Gamma-point only calculations were performed with a plane wave cutoff of 500 Ry. 
We performed GGA calculations with the PBEsol \cite{perdew2008} exchange-correlation functional (shown to provide the best overall agreement with experimental structural and electronic properties\cite{Skelton2015,Giraldo-Gallo2016}) and G\"{o}decker-Teter-Hutter (GTH) pseudopotentials \cite{Hartwigsen1998,Krack2005}. 
We used valence electron configurations $6s^26p^2$ for lead, and $5s^2 5p^4$ for tellurium. 
Spin-orbit coupling was not included. 
Calculations were performed with a $3\times3\times3$ supercell of the conventional (cubic) cell (216 atoms) at temperatures between $50$~K and $500$~K in steps of $50$~K. 
Long simulations ($150$ ps) in the isothermal-isobaric ($NpT$) ensemble using the thermostat developed by \citet{Bussi2007} were performed, followed by $150$ ps production runs in the microcanonical ($NVE$)   ensemble. The timestep used was $2$ fs. 
To check for finite-size effects, 
one run with a $6\times6\times6$ supercell ($1728$ atoms) was performed at $300$ K with a reduced simulation time of $60+60$ ps; this run was also used to analyze in detail the local structure. 

VASP~\cite{kresse1996} calculations were performed using the PAW~\cite{bloechl1994,kresse1999} implementation of density functional theory (DFT). 
We used the PBEsol~\cite{perdew2008} exchange-correlation functional and a plane-wave energy cutoff of $600$ eV. 
For the $15\times1\times1$ supercell we used a $1\times15\times15$ $\Gamma$-centered $k$-point mesh, while for the $3\times1\times1$ supercell we used a $5\times15\times15$ $\Gamma$-centered $k$-point mesh. 
Spin-orbit coupling was not included. 
We used valence electron configurations $5d^{10}6s^26p^2$ for lead and $5s^25p^4$ for tellurium. 
The unit cell volume was fixed to the equilibrium volume (lattice constant of $6.44$~\AA\ to be compared with the experimental lattice constant of $6.43$~\AA~\cite{bozin2010}) obtained with a full structural relaxation. 

{LAMMPS}~\cite{Lammps} MD simulations were performed with a $6\times6\times6$ and a $10\times10\times10$ supercell of the conventional (cubic) cell ($1728$ and $8000$ atoms) at temperatures between $100$~K and $500$~K in steps of $50$~K. 
$150$~ps simulations in the isothermal-isobaric ($NpT$) ensemble using the thermostat developed by~\citet{Bussi2007} and the Berendsen barostat~\cite{Berendsen1984} were performed, followed by $150$~ps production runs in the microcanonical ($NVE$)   ensemble. The timestep used was $2$~fs. 

\subsection{Design and refinement of the 3D-$\Delta$PDF model}
\label{app:3d-delta-pdf}

The 3D-$\Delta$PDF maps were obtained as the Fourier transform of the diffuse diffraction pattern, with the Bragg reflections cut out from the diffraction pattern as follows:
To be sure that the tails from very strong Bragg peaks were eliminated, volumes of $3\times3\times3$ voxels centered at the Bragg positions were set to zero. 
As the strong diffuse scattering maxima coincide with the Bragg peaks it is not possible to cleanly separate Bragg from diffuse scattering without having additional information available. 
The impact of cutting the Bragg peaks and diffuse scattering at the same time was shown to only significantly impact the long 3D-$\Delta$PDF vectors, while densities at short vectors are almost unaffected by this procedure\cite{Kobas2005a,Kobas2005b}. 
In contrast to Refs.~\onlinecite{Kobas2005a,Kobas2005b} we did not fill the punched Bragg regions with the average of the surrounding, but with zero values. As a consequence some artifacts are present at larger PDF vectors, where the 3D-$\Delta$PDF densities are expected to become very weak. 
In the least squares 3D-$\Delta$PDF refinements, the cut-out volumes were ignored by giving them zero weights.

\paragraph*{Theory.}
The local structure properties of PbTe were refined with the 3D-$\Delta$PDF\cite{Weber2012} modeling program YELL\cite{Simonov2014}. 
In the following we summarize the fundamentals of the 3D-$\Delta$PDF theory of displacive disorder. 

In the harmonic approximation, the diffuse scattering of a compound showing displacive disorder may be calculated as: 
\begin{widetext}
\begin{eqnarray} \label{eq:eq1}
 I_{dif}(\mathbf{h})  =
 \sum_{\mathbf{R}_{uvw}}^{cryst} \sum_{mn}^{cell} \Big[
\exp(-\mathbf{h}^{T}\beta^{mn}_{uvw} \mathbf{h}) -  \exp(-\mathbf{h}^{T}(\beta^{m}_{aver}+\beta^{n}_{aver}) \mathbf{h})\Big]
f_{m}(\mathbf{h}) f_{n}^{*}(\mathbf{h}) \cos [2 \pi \mathbf{h}(\mathbf{R}_{uvw} + \mathbf{r}_{mn})]\,.
\end{eqnarray}
\end{widetext}
The summations run over all atomic pairs with significantly correlated displacements. 
$\mathbf{R}_{uvw}$ is the lattice vector between the unit cells hosting the correlated atoms $m$ and $n$ and $\mathbf{r}_{mn}$ is the average distance between the sites $n$ and $m$ within a unit cell, i.e.\  $\mathbf{R}_{uvw}+\mathbf{r}_{mn}$ is the average vector between atoms $m$ and $n$. 
The average ADPs $\beta^{m}_{aver}$ and $\beta^{n}_{aver}$ as well as $\mathbf{r}_{mn}$ are taken from the  average structure and $f_m$ and $f_n$ are the conventional atomic form factors. 
The structural information about displacive correlations is stored in the $3\times3$ matrix $\beta^{mn}_{uvw}$, whose elements are defined as 
\begin{equation}
\begin{split}
\beta_{uvw,ij}^{mn} &= \langle(u_i^n-u_i^m )(u_j^n-u_j^m ) \rangle_{uvw} \\
&=\beta_{aver,ij}^m+\beta_{aver,ij}^n-2\langle u_i^n u_j^m \rangle_{uvw}
\end{split}
\end{equation}
where  $u_i^n$  is the $i$-th vector component of the displacement of atom $n$ from its average position in fractional units and $\langle \ldots \rangle_{uvw}$ indicates  space and time averaging of joint displacements of all atoms $m$ and $n$ that are $uvw$ unit cells apart.  The displacement covariances 
\begin{equation}
Cov_{uvw,ij}^{mn}=\langle u_i^m u_j^n \rangle_{uvw}
\end{equation}
are the structural variables that are optimized in YELL. 
For a more convenient comparison of the joint displacements of atoms we define the correlation matrix $Cor_{uvw}^{mn}$ with the matrix elements 
\begin{equation}
Cor_{uvw,ij}^{mn}=\frac{\langle u_i^m u_j^n \rangle_{uvw}}{\sqrt{\beta_{aver,ii}^n\beta_{aver,jj}^m}}\,,
\end{equation}
whose numerical values may range between $-1$ and $1$. The extreme values represent perfect anti- and in-phase correlations, respectively. 

\paragraph*{Defining the model.}

For modeling the real structure we refer to the harmonic average structure displacement model, because it implies no arbitrary constraint such as $U_{iso}^{\text{Pb}}$ = $U_{iso}^{\text{Te}}$. 
Furthermore, the probability density functions of the harmonic and split models are very similar and the choice of the average structure model is not expected to have a major impact on the extracted information.
\begin{figure}[hptb]
	\includegraphics[width=.95\columnwidth]{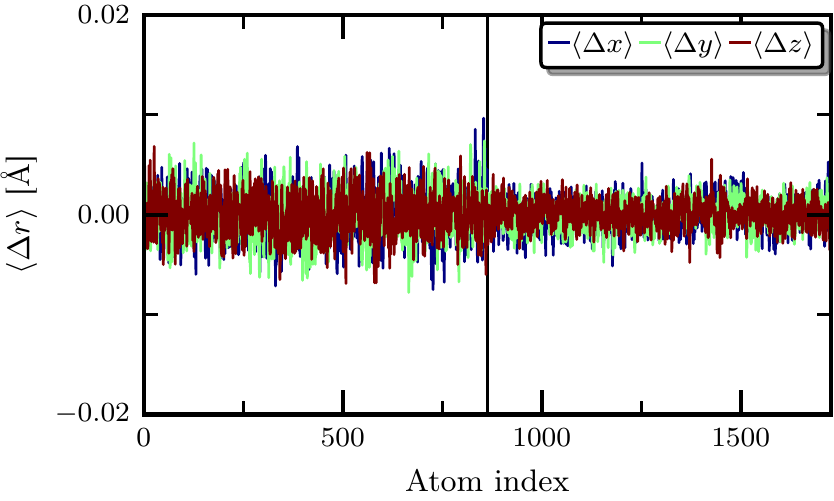}
	\caption{Average displacements of all atoms from the rocksalt high-symmetry positions during the MD simulation at 300 K. The left side is for Pb atoms, the right for Te atoms.}
	\label{fig:app-atom-displ}
\end{figure}

Most of the interatomic vectors in PbTe are nicely resolved in the 3D-$\Delta$PDF maps, with the only exception being that for each Pb/Pb pair there is one Te/Te pair having exactly the same average interatomic vector. Such pair correlation parameters were constrained to the same values. 
The 3D-$\Delta$PDF maps show significant correlations up to about 100 \AA. Refinement of all symmetry-independent coefficients would require a least-squares optimization of about 3500 independent parameters - a task that would be  beyond our computational capabilities. 
To overcome this problem we identified dependencies among the pair correlation parameters, finding by trial-and-error that homo- and heteroatomic pair correlations decay exponentially along $\langle100\rangle$ directions for all except some very short interatomic vectors. The final model was as follows:  
the covariance matrices  $Cov_{uvw}^{mn}$ of the pairs corresponding to interatomic vectors  $\langle 0.5 0 0\rangle$, $\langle 1 0 0\rangle$, $\langle 1.5 0 0\rangle$, $\langle 2 0 0\rangle$, $\langle 2.5 0 0\rangle$ and $\langle 3 0 0\rangle$ were refined independently. 
For any other pairs in the asymmetric unit of the point group $m\bar{3}m$ (i.e. $x \geq y \geq z \geq 0$) the covariance parameters were constrained to an exponential decay according to the relation $Cov_{uvw,ij}^{mn} (xyz)=a_{uvw,ij}^{mn} exp(-b_{uvw,ij}^{mn} x)$, where $a_{uvw,ij}^{mn}$ and $b_{uvw,ij}^{mn}$ were refined separately for sequences of homo- and heteroatomic pairs.
\begin{figure}[hptb]%
	\includegraphics[width=.85\columnwidth]{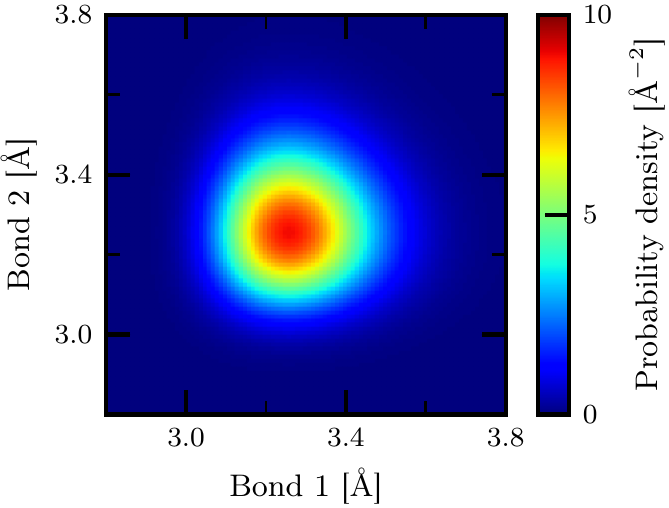}
	\caption{Probability density for the reference state describing uncorrelated bonds; all pairs of bonds at least 15 \AA\ apart were considered.
		}
	\label{fig:app-corrs4-neut}
\end{figure}%
\begin{figure}[hptb]%
	\includegraphics[width=.95\columnwidth]{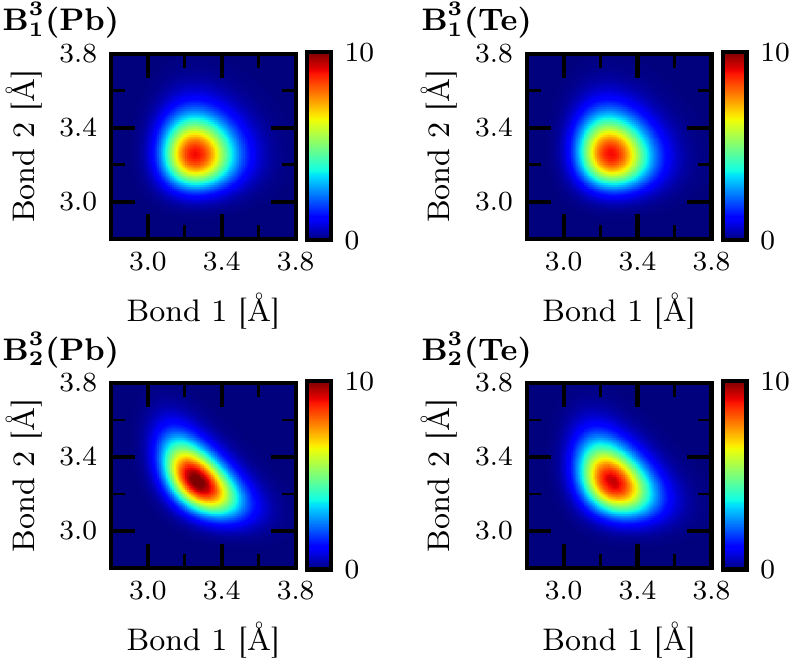}
	\caption{Probability densities of three-body correlations. The color scale represents the probability density in \AA$^{-2}$. 
	}
	\label{fig:app-corrs3}
\end{figure}%

The exponential decay was assumed to be along the main crystallographic axes, because the 3D-$\Delta$PDF maps clearly indicate that the correlations are strongest along such directions. 
In the case of very weak signals it was not possible to refine $a_{uvw,ij}^{mn}$ and $b_{uvw,ij}^{mn}$ independently without obtaining strong numerical correlations. 
Reasonable fits were then obtained with the constraint $b_{uvw,ij}^{mn}=1$, which was found by trial-and-error, and only $a_{uvw,ij}^{mn}$ was refined. The Laue symmetry  $m\bar{3}m$ was applied to all pair correlations. 
The final model comprised $363$ independent parameters to be optimized against the diffuse diffraction data at each temperature. 
Even with this reduced number of parameters, computer memory limitations did not allow refinement of all parameters at the same time, given the large number of voxels in the refinement. 
The models were therefore refined in blocks of about 30 parameters and the process was repeated until no further improvements could be observed. A single least-squares run took about one to two hours on a modern desktop computer. 
In total a full refinement took about two working days per temperature point. The results of the refinements are shown in Figs.~\ref{int_obs_calc} and \ref{obs_calc_3DPDF}. 
In general the agreement between the observed and refined intensities or PDF densities is very good and the R-values are very low given the weak diffuse intensities [R(125 K) = 0.20, R(150 K) = 0.19, R(200 K) = 0.17 R(250 K) = 0.15, R(293 K) = 0.14; here R-values are defined as  $R=\frac{\sum_{i}(I_{obs,i}-I_{ref,i})^2}{\sum_{i}I_{obs,i}^2}$, where the summations run over all $i$ voxels that were observed and not cut-out]. 
The increase of R-values at lower temperatures might be explained by the decreasing integral diffuse scattering intensities, which lead to lower signal-to-noise levels in the experimental diffuse data. 
Major disagreements between observed and refined $\Delta$PDF densities are found along the $\langle 100\rangle$ directions, probably due to anharmonic contributions in the pair correlation functions that are not covered by our harmonic 3D-$\Delta$PDF model. 
The anharmonicities are extended over long distances and increase as a function of temperature, consistent with the observations of Ref.~\onlinecite{bozin2010}.  As a consequence of computational and experimental limitations the achievable PDF space resolution is much lower in the single crystal cases as compared to powder PDF studies. Thus, a more detailed analysis of the anharmonic effects is not possible in the present case.  
The observation of long correlation lengths of anharmonic displacements clearly supports the interpretation of Ref.~\onlinecite{bozin2010} that this anharmonicity is not just a consequence of asymmetric pair potentials, but reflects some collective structural distortion over long distances. 
The anharmonic pair correlations are also seen in the $I_{obs} - I_{ref}$ maps, where significant asymmetries along $h00$ directions with respect to integer $h$ positions are observable. A size-effect like distortion, which is frequently made responsible for such asymmetries, can be excluded because of the absence of substitutional disorder.

\vspace{0.5cm}

\section{Comparison of the obtained pair correlations}

Table \ref{table:pair-corrs-cf} presents an overview of our refined and calculated pair correlations for distances up to two unit cells along each direction. 
Note the good agreement between the 3D-$\Delta$PDF and the MD values. 


\section{Average atomic displacements from the MD simulations}

Fig.~\ref{fig:app-atom-displ} shows the average displacement of the atoms from the rocksalt positions during the simulation at $300$ K. 
Consistent with previous MD simulations they are negligibly small.

\section{Higher-order correlations}

In this section we present probability densities for the higher-order correlations that were not shown in the main text. 
We start by showing in Fig.~\ref{fig:app-corrs4-neut} the reference state, which represents the uncorrelated bonds; for its computation all pairs of bonds at least 15 \AA\ apart were considered. 
One can see that the most probable bond length can be found at $a/2$. Note also the asymmetry in the distribution with a longer tail in the direction of longer bonds, consistent with the asymmetry of the nearest-neighbor PDF peak. In fact, by integrating out one of the bonds we directly obtain the radial distribution function (RDF) $R(r)$. 

Next, in Fig.~\ref{fig:app-corrs3} we show the probability densities for the three-body correlations. 
The dependence on the type of shared atom can be clearly recognized. Note that the negative correlation of $B_2^3$ bonds can be observed already by visual inspection. 
This is not the case for more distant four-body correlations. 

\begin{figure*}[hptb]%
	\includegraphics[width=.95\textwidth]{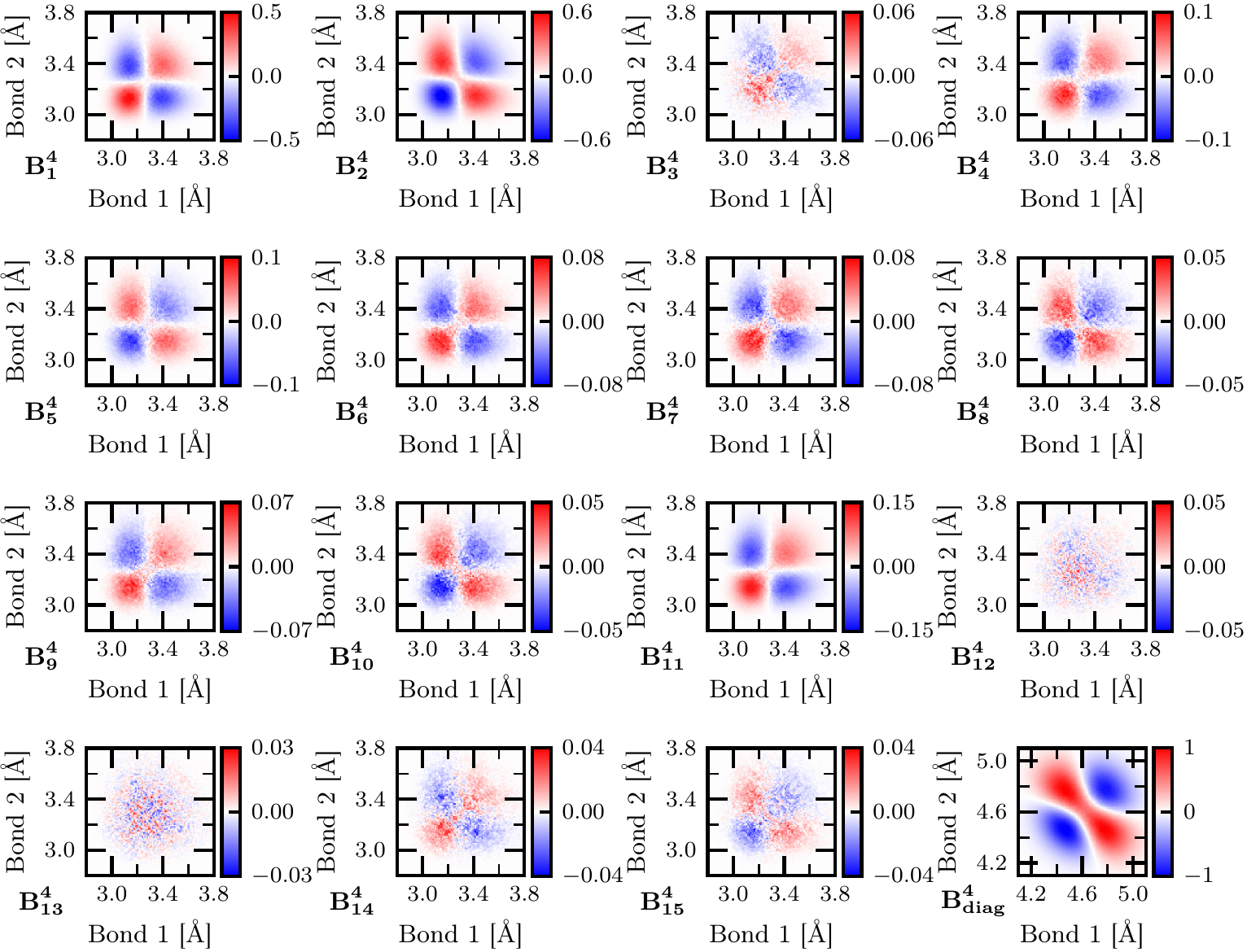}
	\caption{Difference probability densities with respect to the neutral reference of four-body correlations. The color scale represents the probability density in \AA$^{-2}$. Note the different color scale in the panels showing the decay in correlation strength. 
	}
	\label{fig:corrs4}
\end{figure*}

Finally, Fig.~\ref{fig:corrs4} presents all computed four-body correlations. 
Note the different color scale in the different panels showing the decay in correlation strength. 

\begin{table*}
\caption{Pair correlation coefficients $Cor_{uvw,ij}$ within the asymmetric unit 
of the 3D-PDF space up to a maximum distance of two unit cells along each direction; $\Delta \mathbf{x}=(u,\,v,\,w)$ is the interatomic vector in fractional units. 
The 3D-$\Delta$PDF values (labeled as Exp.) were obtained at room temperature, while the MD values at 300~K. 
In the experimental results Pb/Pb and Te/Te correlations cannot be distinguished
but are subject of a combined refinement as described in Appendix \ref{app:3d-delta-pdf}. 
The definition of the correlation coefficients is also described there. 
}
\label{table:pair-corrs-cf}
\begin{tabularx}{\textwidth}{ X X X X X X X X X }
\hline\hline
 $\Delta \mathbf{x}$ & Pairs & Source & $Cor_{11}$ & $Cor_{22}$ & $Cor_{33}$ & $Cor_{12}$ & $Cor_{13}$ & $Cor_{23}$ \tabularnewline\hline
\multirow{3}{*}{0.0  0.0  0.0} & Homo & Exp. & 1.0000 & 1.0000 & 1.0000 & 0.0000 & 0.0000 & 0.0000 \tabularnewline 
              & PbPb & MD   & 1.0000 & 1.0000 & 1.0000 & 0.0000 & 0.0000 & 0.0000 \tabularnewline 
              & TeTe & MD   & 1.0000 & 1.0000 & 1.0000 & 0.0000 & 0.0000 & 0.0000 \tabularnewline \hline
\multirow{2}{*}{0.5  0.0  0.0} & \multirow{2}{*}{PbTe} & Exp. & 0.4879 & 0.1602 & 0.1602 & 0.0000 & 0.0000 & 0.0000 \tabularnewline 
              &      & MD   & 0.4820 & 0.1252 & 0.1289 & 0.0001 & 0.0001 & 0.0068 \tabularnewline \hline 
\multirow{3}{*}{0.5  0.5  0.0} & Homo & Exp. & 0.1552 & 0.1552 & 0.0812 & 0.0581 & 0.0000 & 0.0000 \tabularnewline 
              & PbPb & MD   & 0.1101 & 0.1080 & 0.0462 & 0.0605 & 0.0012 & 0.0022 \tabularnewline 
              & TeTe & MD   & 0.1409 & 0.1388 & 0.0643 & 0.0206 & 0.0009 & 0.0004 \tabularnewline \hline
\multirow{2}{*}{0.5  0.5  0.5} & \multirow{2}{*}{PbTe} & Exp. & 0.0853 & 0.0853 & 0.0853 & 0.0024 & 0.0024 & 0.0024 \tabularnewline 
              &      & MD   & 0.0420 & 0.0728 & 0.0458 &-0.0001 & 0.0002 & 0.0006 \tabularnewline \hline
\multirow{3}{*}{1.0  0.0  0.0} & Homo & Exp. & 0.4051 & 0.0865 & 0.0865 & 0.0000 & 0.0000 & 0.0000 \tabularnewline 
              & PbPb & MD   & 0.3872 & 0.0560 & 0.0556 &-0.0013 &-0.0004 & 0.0021 \tabularnewline 
              & TeTe & MD   & 0.3780 & 0.0561 & 0.0554 & 0.0002 & 0.0010 & 0.0069 \tabularnewline \hline
\multirow{2}{*}{1.0  0.5  0.0} & \multirow{2}{*}{PbTe} & Exp. & 0.1255 & 0.0755 & 0.0539 & 0.0111 & 0.0000 & 0.0000 \tabularnewline 
              &      & MD   & 0.0996 & 0.0504 & 0.0234 & 0.0104 &-0.0000 & 0.0001 \tabularnewline \hline
\multirow{3}{*}{1.0  0.5  0.5} & Homo & Exp. & 0.0778 & 0.0499 & 0.0499 & 0.0060 & 0.0060 & 0.0074 \tabularnewline 
              & PbPb & MD   & 0.0441 & 0.0217 & 0.0161 & 0.0031 & 0.0046 & 0.0064 \tabularnewline 
              & TeTe & MD   & 0.0625 & 0.0300 & 0.0263 & 0.0020 & 0.0013 & 0.0013 \tabularnewline \hline 
\multirow{3}{*}{1.0  1.0  0.0} & Homo & Exp. & 0.0791 & 0.0791 & 0.0348 & 0.0116 & 0.0000 & 0.0000 \tabularnewline 
              & PbPb & MD   & 0.0553 & 0.0488 & 0.0054 & 0.0084 &-0.0011 & 0.0016 \tabularnewline 
              & TeTe & MD   & 0.0513 & 0.0442 & 0.0104 & 0.0061 &-0.0004 & 0.0021 \tabularnewline \hline
\multirow{2}{*}{1.0  1.0  0.5} & \multirow{2}{*}{PbTe} & Exp. & 0.0526 & 0.0526 & 0.0384 & 0.0041 & 0.0024 & 0.0024 \tabularnewline 
              &      & MD   & 0.0238 & 0.0204 & 0.0093 & 0.0017 & 0.0006 & 0.0007 \tabularnewline \hline
\multirow{3}{*}{1.0  1.0  1.0} & Homo & Exp. & 0.0378 & 0.0378 & 0.0378 & 0.0038 & 0.0038 & 0.0038 \tabularnewline 
              & PbPb & MD   & 0.0063 & 0.0207 &-0.0013 & 0.0021 &-0.0014 & 0.0014 \tabularnewline 
              & TeTe & MD   & 0.0090 & 0.0248 & 0.0015 & 0.0011 &-0.0011 & 0.0011 \tabularnewline \hline
\multirow{2}{*}{1.5  0.0  0.0} & \multirow{2}{*}{PbTe} & Exp. & 0.2145 & 0.0392 & 0.0392 & 0.0000 & 0.0000 & 0.0000 \tabularnewline 
              &      & MD   & 0.2068 & 0.0118 & 0.0098 & 0.0000 &-0.0002 &-0.0018 \tabularnewline \hline
\multirow{3}{*}{1.5  0.5  0.0} & Homo & Exp. & 0.1079 & 0.0390 & 0.0299 & 0.0241 & 0.0000 & 0.0000 \tabularnewline 
              & PbPb & MD   & 0.0807 & 0.0094 &-0.0013 & 0.0217 & 0.0004 & 0.0005 \tabularnewline 
              & TeTe & MD   & 0.1053 & 0.0133 & 0.0054 & 0.0096 & 0.0006 & 0.0008 \tabularnewline \hline 
\multirow{2}{*}{1.5  0.5  0.5} & \multirow{2}{*}{PbTe} & Exp. & 0.0664 & 0.0331 & 0.0331 & 0.0018 & 0.0018 &-0.0002 \tabularnewline 
              &      & MD   & 0.0472 & 0.0027 & 0.0027 & 0.0005 & 0.0012 &-0.0008 \tabularnewline \hline
\multirow{2}{*}{1.5  1.0  0.0} & \multirow{2}{*}{PbTe} & Exp. & 0.0640 & 0.0369 & 0.0255 & 0.0078 & 0.0000 & 0.0000 \tabularnewline 
              &      & MD   & 0.0391 & 0.0086 &-0.0026 & 0.0066 & 0.0002 &-0.0000 \tabularnewline \hline 
\multirow{3}{*}{1.5  1.0  0.5} & Homo & Exp. & 0.0478 & 0.0297 & 0.0230 & 0.0052 & 0.0053 & 0.0017 \tabularnewline 
              & PbPb & MD   & 0.0192 & 0.0010 &-0.0032 & 0.0023 & 0.0028 & 0.0004 \tabularnewline 
              & TeTe & MD   & 0.0263 & 0.0045 &-0.0000 & 0.0014 & 0.0006 &-0.0003 \tabularnewline \hline
\multirow{2}{*}{1.5  1.0  1.0} & \multirow{2}{*}{PbTe} & Exp. & 0.0366 & 0.0248 & 0.0248 & 0.0026 & 0.0026 & 0.0010 \tabularnewline 
              &      & MD   & 0.0076 & 0.0001 &-0.0089 & 0.0009 &-0.0002 & 0.0001 \tabularnewline \hline
\multirow{3}{*}{1.5  1.5  0.0} & Homo & Exp. & 0.0358 & 0.0358 & 0.0151 & 0.0150 & 0.0000 & 0.0000 \tabularnewline 
              & PbPb & MD   & 0.0096 & 0.0048 &-0.0053 & 0.0133 & 0.0011 & 0.0024 \tabularnewline 
              & TeTe & MD   & 0.0113 & 0.0070 &-0.0054 & 0.0064 & 0.0010 & 0.0013 \tabularnewline \hline
\multirow{2}{*}{1.5  1.5  0.5} & \multirow{2}{*}{PbTe} & Exp. & 0.0299 & 0.0299 & 0.0172 & 0.0038 & 0.0007 & 0.0007 \tabularnewline 
              &      & MD   & 0.0031 & 0.0001 &-0.0069 & 0.0013 & 0.0001 & 0.0003 \tabularnewline \hline
\multirow{3}{*}{1.5  1.5  1.0} & Homo & Exp. & 0.0228 & 0.0228 & 0.0159 & 0.0052 & 0.0021 & 0.0021 \tabularnewline 
              & PbPb & MD   &-0.0033 &-0.0087 &-0.0081 & 0.0006 & 0.0017 &-0.0005 \tabularnewline 
              & TeTe & MD   &-0.0008 &-0.0045 &-0.0086 &-0.0000 &-0.0001 & 0.0010 \tabularnewline \hline
\multirow{2}{*}{1.5  1.5  1.5} & \multirow{2}{*}{PbTe} & Exp. & 0.0172 & 0.0172 & 0.0172 & 0.0016 & 0.0016 & 0.0016 \tabularnewline 
              &      & MD   &-0.0102 &-0.0075 &-0.0132 & 0.0007 &-0.0005 & 0.0005 \tabularnewline \hline 
\multirow{3}{*}{2.0  0.0  0.0} & Homo & Exp. & 0.1798 & 0.0170 & 0.0170 & 0.0000 & 0.0000 & 0.0000 \tabularnewline 
              & PbPb & MD   & 0.1829 &-0.0041 &-0.0179 & 0.0018 &-0.0003 &-0.0013 \tabularnewline 
              & TeTe & MD   & 0.1710 & 0.0065 &-0.0101 &-0.0001 &-0.0000 &-0.0022 \tabularnewline \hline
\multirow{2}{*}{2.0  0.5  0.0} & \multirow{2}{*}{PbTe} & Exp. & 0.0873 & 0.0278 & 0.0198 & 0.0052 & 0.0000 & 0.0000 \tabularnewline 
              &      & MD   & 0.0706 &-0.0056 &-0.0052 & 0.0038 &-0.0000 &-0.0001 \tabularnewline \hline
\end{tabularx}
\end{table*}

\begin{table*}
\begin{tabularx}{\textwidth}{ X X X X X X X X X }
\hline\hline
 $\Delta \mathbf{x}$ & Pairs & Source & $Cor_{11}$ & $Cor_{22}$ & $Cor_{33}$ & $Cor_{12}$ & $Cor_{13}$ & $Cor_{23}$ \tabularnewline\hline
\multirow{3}{*}{2.0  0.5  0.5} & Homo & Exp. & 0.0591 & 0.0184 & 0.0184 & 0.0040 & 0.0040 & 0.0027 \tabularnewline 
              & PbPb & MD   & 0.0376 &-0.0008 &-0.0125 & 0.0021 & 0.0021 & 0.0006 \tabularnewline 
              & TeTe & MD   & 0.0568 & 0.0033 &-0.0092 & 0.0015 & 0.0020 & 0.0012 \tabularnewline \hline
\multirow{3}{*}{2.0  1.0  0.0} & Homo & Exp. & 0.0585 & 0.0215 & 0.0128 & 0.0073 & 0.0000 & 0.0000 \tabularnewline 
              & PbPb & MD   & 0.0365 &-0.0041 &-0.0072 & 0.0062 & 0.0004 & 0.0002 \tabularnewline 
              & TeTe & MD   & 0.0379 &-0.0030 &-0.0061 & 0.0036 & 0.0000 & 0.0006 \tabularnewline \hline
\multirow{2}{*}{2.0  1.0  0.5} & \multirow{2}{*}{PbTe} & Exp. & 0.0420 & 0.0212 & 0.0141 & 0.0028 & 0.0015 & 0.0006 \tabularnewline 
              &      & MD   & 0.0193 &-0.0057 &-0.0073 & 0.0013 & 0.0003 &-0.0001 \tabularnewline \hline
\multirow{3}{*}{2.0  1.0  1.0} & Homo & Exp. & 0.0308 & 0.0156 & 0.0156 & 0.0027 & 0.0027 & 0.0011 \tabularnewline 
              & PbPb & MD   & 0.0070 &-0.0021 &-0.0132 & 0.0017 & 0.0010 & 0.0002 \tabularnewline 
              & TeTe & MD   & 0.0100 &-0.0029 &-0.0133 & 0.0001 &-0.0011 & 0.0013 \tabularnewline \hline
\multirow{2}{*}{2.0  1.5  0.0} & \multirow{2}{*}{PbTe} & Exp. & 0.0329 & 0.0202 & 0.0114 & 0.0054 & 0.0000 & 0.0000 \tabularnewline 
              &      & MD   & 0.0040 &-0.0046 &-0.0109 & 0.0045 & 0.0001 &-0.0002 \tabularnewline \hline
\multirow{3}{*}{2.0  1.5  0.5} & Homo & Exp. & 0.0274 & 0.0168 & 0.0099 & 0.0050 & 0.0016 & 0.0016 \tabularnewline 
              & PbPb & MD   &-0.0010 &-0.0053 &-0.0103 & 0.0023 & 0.0018 & 0.0019 \tabularnewline 
              & TeTe & MD   & 0.0020 &-0.0047 &-0.0125 & 0.0013 & 0.0004 & 0.0004 \tabularnewline \hline
\multirow{2}{*}{2.0  1.5  1.0} & \multirow{2}{*}{PbTe} & Exp. & 0.0230 & 0.0152 & 0.0113 & 0.0025 & 0.0015 & 0.0008 \tabularnewline 
              &      & MD   &-0.0066 &-0.0102 &-0.0124 & 0.0003 & 0.0005 & 0.0008 \tabularnewline \hline
\multirow{3}{*}{2.0  1.5  1.5} & Homo & Exp. & 0.0163 & 0.0104 & 0.0104 & 0.0023 & 0.0023 & 0.0020 \tabularnewline 
              & PbPb & MD   &-0.0117 &-0.0107 &-0.0131 & 0.0006 & 0.0012 & 0.0020 \tabularnewline 
              & TeTe & MD   &-0.0121 &-0.0121 &-0.0160 & 0.0006 &-0.0002 & 0.0002 \tabularnewline \hline
\multirow{3}{*}{2.0  2.0  0.0} & Homo & Exp. & 0.0211 & 0.0211 & 0.0062 & 0.0081 & 0.0000 & 0.0000 \tabularnewline 
              & PbPb & MD   &-0.0043 &-0.0032 &-0.0226 & 0.0073 &-0.0039 &-0.0033 \tabularnewline 
              & TeTe & MD   &-0.0068 &-0.0050 &-0.0217 & 0.0022 & 0.0004 & 0.0004 \tabularnewline \hline
\multirow{2}{*}{2.0  2.0  0.5} & \multirow{2}{*}{PbTe} & Exp. & 0.0187 & 0.0187 & 0.0076 & 0.0042 & 0.0005 & 0.0005 \tabularnewline 
              &      & MD   &-0.0078 &-0.0074 &-0.0208 & 0.0015 & 0.0007 & 0.0004 \tabularnewline \hline
\multirow{3}{*}{2.0  2.0  1.0} & Homo & Exp. & 0.0150 & 0.0150 & 0.0069 & 0.0034 & 0.0012 & 0.0012 \tabularnewline 
              & PbPb & MD   &-0.0111 &-0.0109 &-0.0202 & 0.0004 & 0.0014 & 0.0007 \tabularnewline 
              & TeTe & MD   &-0.0110 &-0.0110 &-0.0203 & 0.0001 & 0.0009 & 0.0009 \tabularnewline \hline
\multirow{2}{*}{2.0  2.0  1.5} & \multirow{2}{*}{PbTe} & Exp. & 0.0121 & 0.0121 & 0.0079 & 0.0017 & 0.0011 & 0.0011 \tabularnewline 
              &      & MD   &-0.0134 &-0.0145 &-0.0206 & 0.0003 & 0.0006 & 0.0004 \tabularnewline \hline
\multirow{3}{*}{2.0  2.0  2.0} & Homo & Exp. & 0.0079 & 0.0079 & 0.0079 & 0.0012 & 0.0012 & 0.0012 \tabularnewline 
              & PbPb & MD   &-0.0234 &-0.0187 &-0.0163 & 0.0010 &-0.0005 &-0.0010 \tabularnewline 
              & TeTe & MD   &-0.0219 &-0.0231 &-0.0179 & 0.0008 & 0.0002 & 0.0001 \tabularnewline \hline\hline
\end{tabularx}
\end{table*}
\end{document}